\documentclass[aps,prd,showpacs,nofootinbib,superscriptaddress,preprintnumbers,twocolumn]{revtex4-1}
\usepackage[utf8]{inputenc}
\usepackage{graphicx,amsmath,amssymb,soul,bbm,subcaption,dcolumn}
\usepackage{booktabs, multirow,array,float,enumitem}
\usepackage{caption}
\newcolumntype{C}{>{$}c<{$}}
\AtBeginDocument{
\heavyrulewidth=.08em
\lightrulewidth=.05em
\cmidrulewidth=.03em
\belowrulesep=.65ex
\belowbottomsep=0pt
\aboverulesep=.4ex
\abovetopsep=0pt
\cmidrulesep=\doublerulesep
\cmidrulekern=.5em
\defaultaddspace=.5em
}
\usepackage[dvipsnames]{xcolor}
\usepackage{lmodern,dsfont}
\newcommand{\be}{\begin{eqnarray}}
\newcommand{\ee}{\end{eqnarray}}

\def\fm {\mathop{\hbox{fm}}}
\def\MeV {\mathop{\hbox{MeV}}}

\def\Re {\mathop{\hbox{Re}}}
\def\Im {\mathop{\hbox{Im}}}
\def\Tr {\mathop{\hbox{Tr}}}

\usepackage{amsmath} 
\usepackage{dsfont}  
\usepackage{bm}      

\def\beq{\begin{equation}}
\def\eeq{\end{equation}}
\def\beqs#1\eeqs{\beq\begin{split} #1 \end{split}\eeq}

\def\comment#1{}

\def\av#1{ \left\langle #1 \right\rangle }

\usepackage{microtype}

\usepackage[colorlinks=true,backref=false, linktocpage=true,
citecolor=PineGreen,urlcolor=PineGreen,linkcolor=PineGreen,pdfpagemode=UseOutlines]{hyperref}

\hypersetup{%
  bookmarksnumbered=true,
  pdftitle = {},
  pdfsubject = {},
  pdfauthor = {},
  pdfkeywords = {}
}

\begin{document}
\title{Extraction of isoscalar $\pi\pi$ phase-shifts from lattice QCD}

\author{D.\ Guo}
\email{guodehua@gwmail.gwu.edu}
\affiliation{The George Washington University, Washington, DC 20052, USA}

\author{A.\ Alexandru}
\email{aalexan@gwu.edu}
\affiliation{The George Washington University, Washington, DC 20052, USA}
\affiliation{Department of Physics, University of Maryland, College Park, MD 20742, USA}
\affiliation{Albert Einstein Center for Fundamental Physics, Institute for Theoretical Physics, University of Bern, Sidlerstrasse 5, CH-3012 Bern, Switzerland}

\author{R.\ Molina}
\email{ramope@if.usp.br}
\affiliation{Institute of Physics of the University of S\~ao Paulo, Rua do Mat\~ao, 1371, Butant\~a, S\~ao Paulo, 05508-090, Brazil}

\author{M.\ Mai}
\email{maximmai@gwu.edu}
\affiliation{The George Washington University, Washington, DC 20052, USA}

\author{M.\ D\"oring}
\email{doring@gwu.edu}
\affiliation{The George Washington University, Washington, DC 20052, USA}
\affiliation{Thomas Jefferson National Accelerator Facility, Newport News, VA 23606, USA}


\begin{abstract}
We conduct a two-flavor ($N_f=2$) lattice QCD calculation of the elastic phase-shifts for pion-pion scattering in the scalar, isoscalar channel (the $\sigma$-meson). The calculation is performed for two quark masses corresponding to a pion mass of $315\MeV$ and $227\MeV$. The $\sigma$-meson parameters are extracted using various parametrizations of the scattering amplitude. The results obtained from a chiral unitary parametrization are extrapolated to the physical point and read ${M_\sigma = (440^{+10}_{-16}(50) - i\,240(20)(25))\MeV}$, where the uncertainties in the parentheses denote the stochastic and systematic ones. The behavior of the $\sigma$-meson parameters with increasing pion mass is discussed as well.
\end{abstract}

\maketitle

\section{Introduction}

The lightest excited state in the spectrum of hadrons is at the same time one of the most controversial. As described in detail in an extensive review~\cite{Pelaez:2015qba}, its properties (mass and width) and even presence were debated for a long time. Many precise analyses finally led to the currently accepted ranges for mass, $400-550$~MeV, and width, $400-700$~MeV, of the so-called $\sigma$ or $f_0(500)$ resonance~\cite{Patrignani:2016xqp}. This scalar/isocalar excited state has a dominant decay channel to two pions, with a very uncommon shape of the partial wave in this channel.

Lattice QCD is the only method to compute the hadron properties directly in terms of quark-gluon QCD dynamics. In the context of $\pi\pi$ scattering many important results have been reported in the $I=L=1$ channel~\cite{Aoki:2007rd, Lang:2011mn, Pelissier:2011ib, Pelissier:2012pi,Guo:2015dde,Bali:2015gji, Guo:2016zos, Aoki:2011yj, Bulava:2016mks, Dudek:2012xn,Wilson:2015dqa}. Due to the presence of disconnected diagrams the corresponding calculations in the $I=L=0$ channel were not possible for a long time, despite early pioneering works~\cite{Alford:2000mm, Prelovsek:2010kg, Fu:2013ffa}. The first results have been reported recently by the Hadron Spectrum Collaboration~\cite{Briceno:2016mjc}, extracting also phase-shifts using the L\"uscher framework~\cite{Luscher:1990ux} in combination with moving frames~\cite{Rummukainen:1995vs}. The data have then been analyzed and extrapolated to the physical point using the Inverse Amplitude Method in Ref.~\cite{Doring:2016bdr}. The isoscalar scattering length at three pion masses has been extracted recently by the European Twisted Mass collaboration~\cite{Liu:2016cba}.

In the present work we report new $N_f=2$ lattice QCD results for two different pion masses ($M_\pi=227$ and $M_\pi=315$~MeV) and analyze them.  To extract a robust energy spectrum in a specific scattering channel, a large and proper interpolating field basis is required. To have an interpolating field basis that has the correct quantum number and symmetry as the scattering channel and has enough overlap with the relevant eigenstates of the system, both quark-antiquark ($q\bar{q}$ operators) and two-hadron interpolators are included. We perform a variational analysis~\cite{Luscher:1990ck} with this interpolating field basis to extract several low lying energy states that are in the elastic scattering region. The evaluation of the correlation functions are carried out using the Laplacian-Heaviside smearing method~\cite{Peardon:2009gh,Morningstar:2011ka} for all the possible quark diagrams calculated from the Wick-contraction procedure. To have more phase-shift data points in different kinematic region so as to better describe the energy dependence of the $\pi\pi$ scattering phase-shifts, we implement our calculation in three boxes with different elongation factor and two total-momentum frames, one being the rest frame $\bm P=(0,0,0)$, the other one moving along the elongated direction with $\bm P=(0,0,1)$, the smallest non-zero momentum allowed by the boundary conditions. Since we are interested in two-particle scattering in the elastic region, the physical observables such as the phase-shift can be obtained from the energy spectrum in finite volume using the L\"{u}scher formula~\cite{Luscher:1990ux}.

In the second step of the present work we perform an energy-dependent analysis using directly the set of energy eigenvalues at different elongations and momenta. This allows to take into account the correlations between different energy eigenvalues. To this end, we formulate the scattering amplitude in a similar manner to the $K$-matrix approach, which allows to access finite-volume energy eigenlevels (as positions of the poles of this amplitude). The free parameters of the parameterizations are fitted to reproduce the lattice data and then used in the infinite-volume formulation to determine the parameters of the isoscalar $\pi\pi$ resonance. We use different parametrizations to estimate the systematic uncertainty from this source. Specifically, we consider a general expansion in an energy-variable conformally  mapping the energy plane to the unit disk, similar to the analysis of Refs.~\cite{Yndurain:2007qm, Caprini:2008fc}. Second, we employ a model based on the chiral unitary approach (UChPT), used, e.g., in Refs.~\cite{Hu:2017wli,Hu:2016shf,Guo:2016zos}.  Subsequently, the UChPT amplitude is extrapolated to the physical point. Our final result, based on all lattice data presented here with and without the isovector channel data~\cite{Guo:2016zos}, reads $M_\sigma^{\rm phys}=(440^{+10}_{-16}-i\,240_{-20}^{+20})$~MeV and agrees with the result of the  most recent analysis of experimental data~\cite{Pelaez:2015qba} within the quoted $1\sigma$ region. Additionally, the study of the pion mass dependence of resonance mass and coupling to the $\pi\pi$ channel is studied in a broader range of $M_\pi$, as it was done in Refs.~\cite{Doring:2016bdr,Hanhart:2008mx}.

The paper is organized as following: In Sec.~\ref{sec:lattice} we describe the details of the lattice calculation to obtain the energy eigenvalues and correlation matrices; In Sec.~\ref{sec:analysis} we describe the scattering amplitudes used for the extrapolation of the lattice results in energy and pion mass; The results of these analyses are discussed in Sec.~\ref{subsec:discussion}, and the overall summary is given in Sec.~\ref{sec:summary}.

\section{TECHNICAL DETAILS}
\label{sec:lattice}
\subsection{Interpolating basis}
\label{sec:interpolating_field}
As we mention in the introduction, the composition of the $\sigma$-meson might include different kinds of components. In order to extract low-lying energy states from the correlation function, we perform a variational analysis. In this study, we are interested in two-pion elastic scattering. To better extract the low lying energy levels in the elastic scattering region, we choose a set of interpolating fields with the same quantum number including the quark-antiquark ($q\bar{q}$) and meson-meson interpolating fields in the variational basis. There are several reasons for using a large variational basis. Firstly, it helps resolve energy states that are nearly degenerated. Secondly, it offers large enough overlap with the eigenstates of the Hamiltonian, which can improve the accuracy and stability of the extracted energy spectrum. 

The correlation matrix is constructed from two-point functions of all combinations of the interpolating fields in the variational basis. The elements of the correlation matrix are 
\beq
C_{ij}(t) = \left<\mathcal{O}_i(t)\mathcal{O}_j^{\dagger}(0)\right>\,.
\eeq
We denote the interpolators as $\mathcal{O}_i$ with $i=1,...,N$ with $N$ being the number of interpolating fields in the basis. The eigenvalues of the correlation matrix can be obtained by solving the generalized eigenvalue problem
\beq
\label{eq:generalized_eigenvalues}
C(t_0)^{-\frac{1}{2}}C(t)C(t_0)^{-\frac{1}{2}}\psi^{(n)}(t,t_0)=
\lambda^{(n)}(t,t_0)\psi^{(n)}(t,t_0)
\eeq
for a particular initial time $t_0$ and for each time slice $t$. The energies of the system are then determined from the long-time behavior of the eigenvalues~\cite{Blossier:2009kd}
\beq
\lambda^{(n)}(t,t_0) \propto e^{-E_n t}[1+\mathcal{O}(e^{-\Delta E_nt})]\,, \ \ \ n=1,...,N\,,
\eeq
where the correction term depends on the energy difference $\Delta E_n = E_{N+1}-E_n$. According to this behavior, for the low-lying energy states, the larger interpolating basis we use, the faster the correction vanishes. However, the benefit from enlarging the interpolating basis decreases because the energy eigenstates get denser in the higher-energy part of the spectrum. Our goal is to choose a interpolating basis that is good enough to capture the energy eigenstates in the elastic scattering energy region. 

In this work, we consider lattices with one spatial direction elongated. The corresponding rotational symmetry group for this elongated box is $D_{4h}$ which is a subgroup of the full rotational symmetry group $SO(3)$. Therefore, the angular momenta that label the irreducible representations (irreps) of $SO(3)$ are split into multiplets related to irreps of the $D_{4h}$ group. The resulting split for the lowest angular momentum multiplets is listed in Table~\ref{tab:D4h}. The $\sigma$-meson has angular momentum $L=0$ and positive parity. The irrep $L=0$ maps to the one-dimensional $A_1^+$ irrep. The lowest state in this channel corresponds to a $\pi$-$\pi$ state for which the two pions decay at rest. Therefore, the lowest state changes very little when varying the elongation of the box. The excited states energy changes as we increase the elongation and we can scan this corresponding energy region.

In order to have more energy values data points in different kinematic region, we implement the boosted frame method. The idea is to boost the whole system with a given momentum $\bm P$ in a certain direction. Due to the relativistic effects, the box along the boosted direction is contracted~\cite{Rummukainen:1995vs}. A generic boost direction of the box will further reduce the original symmetry group to a subgroup which depends on the direction of the boost. In this study, we implement a boost to an elongated box parallel to the elongated direction. In this case, the boost reduces the rotational symmetry group of the cubic box from $O_{h}$ to $D_{4h}$, but it does not change the rotational symmetry group for the elongated box, which is still $D_{4h}$.
 
We note that the states in the $A_1^+$ irrep belong to different irreps of $SO(3)$. The $A_1^+$ irrep couples not only to $L=0$, but also to other higher angular momentum channels such as $L=4$ and so on in the $O_h$ group for the cubic box and $L=2$, $L=4$ and so on in the $D_{4h}$ group for the elongated box. However, to study the $\sigma$-meson, we are interested in a low energy region where the two-pion states with relatively small scattering momentum. In this case, the effect from the $L \ge 2$ channels are small and their contribution can be safely neglected because it is kinematically suppressed through the barrier factor. Indeed, the influence of the $D$-wave in the extraction of the $S$-wave from energy eigenvalues in $A_1^+$ has been estimated in Ref.~\cite{Doring:2012eu} using realistic $S$ and $D$ waves; it was found to be small.

\begin{table}[t]
\begin{tabular}{r c c}
\toprule
$~~\ell~~~~~$ & $O_{h}$ & $D_{4h}$ \\
\hline
~~0~~~~~ & $A_1^+$                                        & $A_1^+$\\
~~1~~~~~ & $F_1^-$                                        & $A_2^-\oplus E^-$\\
~~2~~~~~ & $E^+ \oplus F_2^+$                             & $A_1^+\oplus B_1^+ \oplus B_2^+ \oplus E^+$\\
~~3~~~~~ & $A_2^- \oplus F_1^- \oplus F_2^-$              & $A_2^- \oplus B_1^- \oplus B_2^- \oplus 2 E^-$\\
~~4~~~~~ & $A_1^+ \oplus E^+ \oplus F_1^+ \oplus F_2^+$ & $2A_1^+ \oplus A_2^+ \oplus B_1^+ \oplus B_2^+ \oplus 2E^+$\\
\bottomrule
\end{tabular}
\caption{Resolution of angular momentum in terms of irreps of the $O_h$ and the $D_{4h}$ group.}
\label{tab:D4h}
\end{table}

As a result, we focus on the states in the $A_1^+$ irrep. For the volumes considered in this study, there are only two or three low-lying energy states in the elastic scattering energy region. As mentioned before, we need a basis that has overlaps both with the resonance state (to take into account possible two-quark components of the $\sigma$-resonance) and with the states that have a dominant two-pion content. We include four quark-antiquark interpolators in our basis so that we can see their effects on the energy spectrum. These four quark-antiquark interpolators have the form
\beq
\sigma(\Gamma_i(\bm p),t) = \frac{1}{\sqrt{2}}[\bar{u}(t)\Gamma_i(\bm p) u(t) + \bar{d}(t)\Gamma_i(\bm p) d(t)] \,.
\eeq
The $u(t)$ and $d(t)$ denote the up and down quark on the entire $t$ time slice which is a column vector of size $N = 12\times N_x \times N_y \times N_z$. The $\Gamma_i(\bm p)$ represents $N\times N$ matrices. Their form in the creation operators is defined as $\Gamma'_i(\bm p)$ which can be derived using 
\beq
[\sigma(\Gamma_i(\bm p), t)]^{\dagger} = \sigma(\Gamma'_i(\bm p),t) \,.
\eeq
The details for $\Gamma_i(\bm p)$ and $\Gamma'_i(\bm p)$ are listed in the first four row of Table~\ref{tab:interp}. The first interpolator is point-like and the next three interpolators involve $\bar q q$ pairs that are separated by several lattice spacing, defined using the covariant derivative 
\beq
(\nabla_k)_{x,y}^{ab} = U_k^{ab}(x) \delta_{x+\hat k,y} - \delta^{ab}\delta_{x,y} \,.
\eeq
The forth interpolator has a different gamma matrix structure.

In previous studies for the $\rho$-meson resonance in the $\pi$-$\pi$ scattering channel~\cite{Lang:2011mn,Dudek:2012xn,Guo:2016zos}, it was shown that the quark-antiquark interpolators are not sufficient to extract a reliable spectrum in the interacting theory where the actual eigenstates are mixed with quark-antiquark basis states and multi-hadron basis states. The reason is that the quark-antiquark interpolators have little overlap with the multi-hadron states and the overlap is shown to be suppressed by a power of the lattice volume~\cite{Dudek:2012xn}. To solve this problem, we include the pion-pion interpolators in the variational basis. First, we construct the pion-pion interpolators to have isospin $I = 0$ and $I_3 = 0$ which correspond to the $\sigma$-meson:
\beqs
\pi\pi(\bm p_1,\bm p_2) =& \frac{1}{\sqrt{3}}\{\pi^+(\bm p_1) \pi^-(\bm p_2) + \pi^-(\bm p_1) \pi^+(\bm p_2)\\
&+\pi^0 (\bm p_1) \pi^0(\bm p_2) \} \,,
\label{eq:pipi_interpolators}
\eeqs
where $\pi^+$, $\pi^-$ and $\pi^0$ are given by
\beqs
 \\
\pi^+(\bm p,t) &= \sum_{\bm x} \bar{d}(\bm x,t) \gamma_5 u(\bm x,t) e^{i\bm p\bm x}=\bar{d}(t) \Gamma_5(\bm p) u(t) \,, \\
\pi^-(\bm p,t) &= \sum_{\bm x} \bar{u}(\bm x,t) \gamma_5 d(\bm x,t) e^{i\bm p\bm x}=\bar{u}(t) \Gamma_5(\bm p) d(t) \,,
\eeqs
and 
\beqs
\label{eq:pion0_v2}
\pi^0(\bm p,t)&=  \frac{1}{\sqrt{2}}\sum_{\bm x}\{\bar{u}(\bm, x, t)\Gamma_5 e^{i \bm p \bm x}u(\bm x,t) \\
&- \bar{d}(\bm x,t)\Gamma_5 e^{i \bm p \bm x} d(\bm x,t)\} \\
&= \frac{1}{\sqrt{2}}\{ \bar{u}(t) \Gamma_5(\bm p) u(t) - \bar{d}(t) \Gamma_5(\bm p) u(t)\}\,.
\eeqs

{\renewcommand{\arraystretch}{1.5}
\begin{table}[t]
\center
\begin{tabular}{*{5}{>{$}r<{$}}}
\toprule
~~~~~~~~~i~~~~~~~ 
&\phantom{abc}& \Gamma_i(\bm p)~~~~~
&\phantom{ab}& \Gamma_i'(\bm p)~~~~~~~~~~~~ \\
\midrule
~~~~~~~~~1~~~~~~~ 
 && \bm 1 e^{i \bm p}~~~~~                     
 && \bm 1 e^{-i \bm p} ~~~~~~~~~~~~\\
~~~~~~~~~2~~~~~~~ 
 && \nabla_i \bm 1 e^{i\bm p} \nabla_i~~~~~    
 && \nabla_i \bm 1 e^{-i\bm p} \nabla_i~~~~~~~~~~~~\\
~~~~~~~~~3~~~~~~~ 
 && \nabla_i^4 \bm 1 e^{i\bm p} \nabla_i^4~~~~~
 && \nabla_i^4 \bm 1 e^{-i\bm p} \nabla_i^4 ~~~~~~~~~~~~\\
~~~~~~~~~4~~~~~~~ 
 && \gamma_i e^{i\bm p} \nabla_i~~~~~          
 && \gamma_i e^{-i\bm p} \nabla_i ~~~~~~~~~~~~\\
~~~~~~~~~5~~~~~~~ 
 && \gamma_5 e^{i\bm p}~~~~~                   
 && -\gamma_5 e^{-i\bm p} ~~~~~~~~~~~~\\
\bottomrule
\end{tabular}
\caption{Interpolator structure for the quark bilinears for the quark-antiquark interpolators ($i=1-4$) and for pion-pion interpolators ($i=5$).}
\label{tab:interp}
\end{table}
As we mention before, in this study, we focus on the $A_1^+$ irrep which mainly couples to S-wave of the pion-pion scattering. To construct interpolators transforming according to the $A_1^+$ representation, we project the general $\pi\pi$ interpolators into $A_1^+$ using 

\beq
\pi\pi(\bm p_1,\bm p_2)_{A_1^+} = \frac1{|D_{4h}|} \sum_{g\in D_{4h}} \chi_{A_1^+}(g) \pi\pi(R(g) \bm p_1, R(g) \bm p_2) \,,
\eeq
where $R(g)$ implements the rotation associated with the symmetry transformation $g$, and $\chi_{A_1^+}$ is the character of group element $g$ in the $A_1^+$ irrep. 

In order to have more energy levels in different kinematic regions, we implement two different total momenta for the system. The $\pi\pi$ operators in the rest frame $\bm P_0= (0,0,0)$ in $A_1^+$ irrep are as follows:
\beqs
\pi\pi_{000}^{(0)}  &=\pi\pi(\bm p_1 = (0,0,0), \bm p_2 = \bm P_0 - \bm p_1) \,,\\
\pi\pi_{001}^{(0)}  &=\pi\pi(\bm p_1 = (0,0,1), \bm p_2 = \bm P_0 - \bm p_1) \,.
\eeqs
In the boost frame $\bm P_1 = (0,0,1)$, the lowest scattering momentum $\bm p_1 = (0,0,1)$. Therefore, we use the following interpolators
\beqs
\pi\pi_{001}^{(1)}  &=\pi\pi(\bm p_1 = (0,0,1), \bm p_2 = \bm P_i - \bm p_1) \,,\\
\pi\pi_{002}^{(1)}  &=\pi\pi(\bm p_1 = (0,0,2), \bm p_2 = \bm P_i - \bm p_1) \,,\\
\pi\pi_{011}^{(1)}  &= \frac{1}{2} \sum_{\bm p_1 \in {\cal P}} \pi\pi(\bm p_1, \bm p_2 = \bm P_1 - \bm p_1) \,,
\eeqs
where ${\cal P} =\{(0,1,1),(1,0,1),(-1,0,1),(0,-1,1)\}$ are all the possible momenta generated by symmetry transformations $R(g)\bm p$ from $\bm p = (0, 1, 1)$. The value of $\bm p_2$ in the equations above is imposed by momentum conservation. The reason we only apply the positive total momentum $\bm P_1$ in the boosted case is that the expectation values for the correlation functions associated with momentum $\bm P_1$ and $\bm {-P_1}$ are the same due to rotational symmetry. 

With these six interpolators, we construct a $6\times 6$ variational basis. Each entry in the correlation matrix can be calculated through the Wick contraction procedure. There are three types of entries in the correlation matrix,
\beqs
C_{\sigma_i\leftarrow\sigma_j}&= \langle\sigma_i(\bm P, t_f) \sigma_j^\dagger(\bm P, t_i)\rangle\\[2pt]
	&= \av{-[i\bm P f|j'\bm P i] +2\times [i\bm P f][j' \bm P i]}\,,\\
C_{\sigma_i\leftarrow\pi\pi}&=
\av{\sigma_i(\bm P,t_f) \pi\pi(\bm p, \bm P-\bm p, t_i)^\dagger} \\
	&= \sqrt{3/2} \times \big\langle -2[i\bm P f][5'\bm p i|5'\bm P-\bm p i] \\
    &~+ [i\bm P f|5'\bm P-\bm p i|5'\bm p i] \\
    &~+ [i\bm P f|5'\bm p i|5'\bm P-\bm p i]
    \big\rangle \,, \\
C_{\pi\pi\leftarrow\pi\pi}&=\av{\pi\pi(\bm p',\bm P-\bm p',t_f)\pi\pi(\bm p,\bm P-\bm p,t_i)^\dagger}\\
	&=\big\langle
3\times [5\bm p' f|5\bm P-\bm p' f][5' \bm p i| 5'\bm P- \bm p i] \\
	&~+1\times [5\bm p' f|5' \bm p i][5\bm P -\bm p' f|5' \bm P - \bm p i] \\
	&~+1\times [5\bm p' f|5' \bm P -\bm p i] [5\bm P -\bm p'f| 5' \bm p i] \\
	&~-(3/2) \times [5\bm p' f|5 \bm P -\bm p' f| 5'\bm p i| 5' \bm P - \bm p i] \\
    &~-(3/2) \times [5\bm p' f|5' \bm P -\bm p i| 5'\bm p i| 5 \bm P - \bm p' f] \\
	&~-(3/2) \times [5\bm p' f|5\bm P -\bm p'f| 5' \bm P -\bm p i | 5' \bm p i] \\
    &~-(3/2) \times [5\bm p' f|5'\bm p i| 5' \bm P -\bm p i | 5\bm P -\bm p' f] \\
	&~+(1/2)\times [5 \bm p' f| 5' \bm P - \bm p i|5\bm P -\bm p' f | 5' \bm p i]\\
    &~+(1/2)\times [5 \bm p' f| 5' \bm p i|5\bm P -\bm p' f | 5' \bm P-\bm p i]
\big\rangle \,.
\label{eq:contractions_sigma}
\eeqs
The notation above is defined as 
\beq
[i_1\bm p_1 j_1|\ldots|i_k \bm p_k j_k]\equiv 
\Tr\prod_{\alpha=1}^k \Gamma_{i_\alpha}(\bm p_\alpha) M^{-1}(t_{j_\alpha},t_{j_{\alpha+1}}) \,,
\label{eq:trace}
\eeq
where $j_{k+1}$ is defined to be $j_1$ and $M^{-1}(t,t')=\av{u(t)\bar{u}(t')}_F$ is the quark propagator
between time slices $t$ and $t'$ (for more details about the notation
see Ref.~\cite{Guo:2016zos}). 

The calculation of the correlation functions involves the evaluation of the all-to-all propagator which is not practical to compute directly from the position space. We evaluate them in another way using the LapH method. The setup details for the LapH method in this study can be founded in our previous study~\cite{Guo:2016zos}. To calculate the all-to-all propagator projected on
the LapH space requires a large number of fermionic matrix inversions. This was done efficiently using our GPU inverters~\cite{Alexandru:2011ee}.

\subsection{Finite-volume spectrum}
In the $I=0$ channel, each entry of the correlation function contains temporally disconnected diagrams in which the trace only has one time slice. Evaluation of these diagrams is difficult. The reason it that this channel has the same quantum numbers as the vacuum and the correlators do not vanish as the time separation is taken to infinity. The constant contribution has to be subtracted in order to get the expected exponential behavior. In our study, we implement two approaches to subtract the vacuum contribution from temporally disconnected diagrams. In the first method we estimate the vacuum contribution by taking the average of the vacuum bubble and subtract this value from the original correlation functions as 
\beq
\left<\mathcal{O}(t_2) \mathcal{O}^{\dagger}(t_1) \right>_{\text{sub}}
= \left<\mathcal{O}(t_2) \mathcal{O}^{\dagger}(t_1)\right> - \left<\mathcal{O}(t_2)\right> \left< \mathcal{O}^{\dagger}(t_1)\right> \,.
\eeq
The second approach to solve this problem is to consider the so called \textit{shifted correlators} instead of the original correlators in the correlation matrix 
\beq
\label{eq:shift_correlator}
\tilde{C}_{ij}(t)  = C_{ij}(t+d) - C_{ij}(t) \,,
\eeq
where $d$ is the time shift between two correlation functions. In this case, since the vacuum contribution is a constant in the correlator, it is subtracted implicitly when taking the difference of correlation functions. We compare the energy levels  extracted from these two approaches at $M_{\pi}=227\MeV$ with different elongation factor $\eta$ in Fig.~\ref{fig:subtraction_220mev}. 

\begin{figure}
\includegraphics[width=\linewidth]{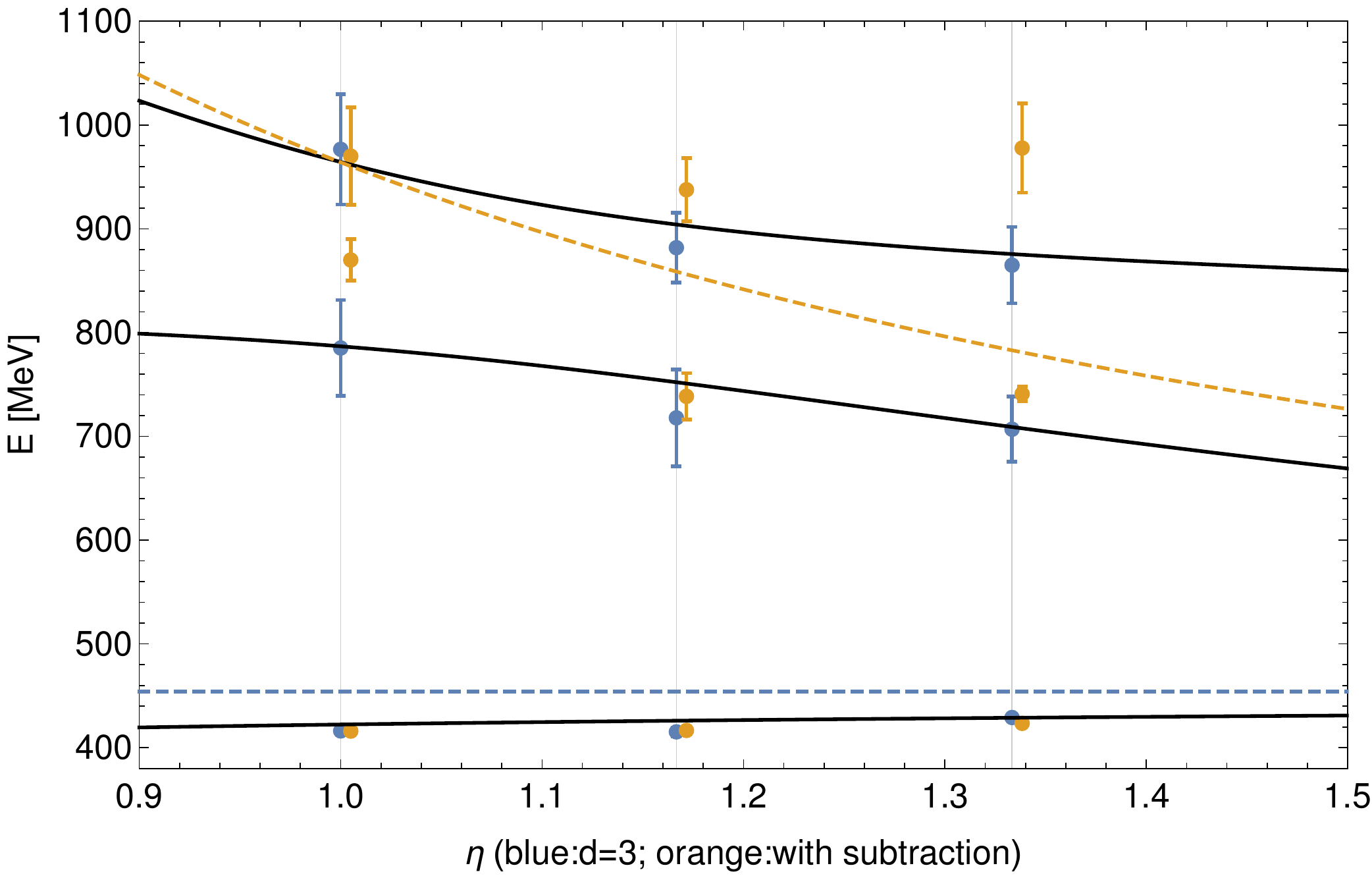}
\caption{
Comparison of the $\sigma$ energy spectrum with direct subtraction (orange) and shifted correlator (blue) methods at pion mass ensembles $m_{\pi}\approx 227\MeV$. Orange points are displaced slightly in the horizontal direction for clarity. The solid curves are a prediction of the UChPT model with a parameter set from the fit to the $\rho$ data~\cite{Guo:2016zos}.}
\label{fig:subtraction_220mev}
\end{figure}

\begin{table*}[t]
\begin{tabular}{@{}*{13}{>{$}c<{$}}@{}}
\toprule
\text{ensemble}& N_t\times N_{x,y}^2\times N_z & \eta & a[\fm]   & N_\text{cfg}  & aM_{\pi}  & am_{N} &am^{pcac}_{u/d}  & af_{\pi}  \\
\midrule
\mathcal{E}_1&48\times24^2\times24  &  1.0  & 0.1210(2)(24) & 300   & 0.1934(5)  & 0.644(6) & 0.01237(9)& 0.0648(8)    \\ 
\mathcal{E}_2&48\times24^2\times30  &  1.25 & -        & -     & -   &    -      &     -     &  -      \\
\mathcal{E}_3&48\times24^2\times48  &  2.0  & -        & -     & -   &   -   &     -     &  -         \\
\mathcal{E}_4&64\times24^2\times24  &  1.0  & 0.1215(3)(24)& 400  & 0.1390(5)  & 0.62(1)& 0.00617(9)& 
0.060(1)  \\
\mathcal{E}_5&64\times24^2\times28  &  1.17 &      -   & -     & -    &     -     &   -       &   -         \\
\mathcal{E}_6&64\times24^2\times32 &  1.33 &      -   & -     & -   &     -      &   -       &   -      \\
\bottomrule
\end{tabular}
\caption{The parameters for the ensembles used in this study. The lattice spacing $a$ for each ensemble is listed as well as the number of gauge configurations $am_{N}$, $af_{\pi}$, and $af_{K}$ represent the nucleon mass, pion decay constant and kaon decay constant in lattice units. The two errors for the lattice spacing are stochastic, from the $w_0/a$ determination, and a systematic one estimated to be 2\%.}
\label{Tab:ensembles}
\end{table*}
We note that the results from direct subtraction have smaller error bars, but they seem to be inconsistent with expectation from the UChPT predictions when using the parameters for this model extracted from the $\rho$ study. Furthermore, these results are also inconsistent with the ones extracted using the moving frame correlators that do not require a vacuum subtraction. It turns out that energy levels extracted using this method are very sensitive to the value of the constant used to subtract the correlator, the vacuum expectation value of the one-point functions generated by the interpolators. One possibility is that the values we computed are biased by wrap-around effects in the time direction. The shifted correlator method does not suffer from this problem, but it generates results with larger error-bars. The reason for this is that, for the same fitting window as in the direct subtraction method, the correlators involved are noisier since the shifted values of the correlators correspond to later times. We decided to use the shifted correlator method for the zero momentum states. For the moving states we do not need any subtraction, since the one-point functions vanish in this case. 

We also looked at the stability of our results with respect to varying the interpolator basis. Our conclusions were similar to the ones derived in our $\rho$ study~\cite{Guo:2016zos}. The only noticeable difference is that for the moving states, we found that the energy levels are more sensitive to the presence of the $\mathcal{O}_4 = \gamma_i \nabla_i$ interpolator in our basis. We believe this is because the other $\bar q q$ operators have the same $\gamma$-matrix structure, all $\gamma=1$, and thus $O_4$ provides a significantly different overlap with the relevant states.

The energy levels extracted for this channel both for zero-momentum and moving states and the details of the fitting parameters are listed in Appendix~\ref{appendix:fitting}.

\subsection{Phase-shift formulas}

As mentioned in the introduction, in this study, we only consider two-pion scattering below the inelastic threshold. To connect the two-hadron state energies determined from lattice QCD with the physical observables, i.e., the phase-shifts in the continuum, we use L\"{u}scher's formula~\cite{Luscher:1990ux} and its extensions to the elongated box~\cite{Feng:2004ua} and to states boosted along the elongated direction~\cite{Lee:2017igf}. Note that the extraction of resonance parameters from the phase-shifts requires a more delicate analysis in case of $\sigma$-resonance. This is due to the fact that the $\sigma$ cannot be described by an ordinary Breit-Wigner shape. Careful implementation of analyticity and unitarity in the scattering amplitude is necessary and will be discussed in the next section.

We illustrate L\"{u}scher's formula for studying $\sigma$-resonance phase-shifts in the $I(J^{PC}) = 0(0^{++})$ scattering channel. The corresponding irrep for the S-wave scattering channel is $A_1^+$ in both the $O_h$ and $D_{4h}$ groups. The $A_1^+$ irrep in $D_{4h}$  couples to angular momenta $J =L=0,2,4...$. Using the argument for the angular momentum cutoff we discussed in section~\ref{sec:interpolating_field}, we assume that the contribution from $J>0$ is negligible. As a result, we  consider the $A_1^+$ irrep under the condition that it is dominated by $J=0$. L\"{u}scher's formula in $A_1^+$ is then given by
\beq
\label{eq:luscher_formula_A1}
\cot \delta_0 = \mathcal{W}_{00} = \frac{\mathcal{Z}_{00}(1,q^2;\eta)}{\pi^{3/2} \eta q}\,.
\eeq
The boost along the elongated direction does not change the symmetry group of the elongated box. Therefore, the boosted version of L\"{u}scher's formula has the same form as Eq.~\eqref{eq:luscher_formula_A1} in $A_1^+$ irrep of the $D_{4h}$ group but with the modification that comes from the boost factor. The detailed derivation can be found in Ref.~\cite{Guo:2016zos}. 

\begin{table*}[th]
\begin{tabular}{clllllcc}
Parametrization~~&Fitted data&\multicolumn{4}{l}{Free parameters}&~~$~~~~n$&~~$\chi_{\rm d.o.f.}^2$\\
\toprule
&&&&&\\[-16pt]
&&~~$B_0$&~~$B_1$ [MeV$^{-1}$]&&
&&\\    
\midrule 
cm1&$\sigma_{227}$
&~~$+11(1)$&~~$+9(3)$&&&~~$~~~~14$&~~$0.7$\\
cm1&$\sigma_{315}$	
&~~$+6(1)$&~~$-2(3)$&&&~~$~~~~15$&~~$0.7$\\ 
\midrule
&&&&&\\[-16pt]
&&~~$B_0$&~~$B_1$ [MeV$^{-1}$]&&
&&\\
\hline
cm2&$\sigma_{227}$
&~~$+11(2)$	&~~$+9(5)$&&&~~$~~~~14$&~~$0.8$\\
cm2&$\sigma_{315}$
&~~$+6(1)$&~~$-3(6)$&&&~~$~~~~15$&~~$0.7$\\ 
\midrule
&&&&&\\[-16pt]
&&$~~L_a$&$~~L_b$&$~~L_c$&&&\\
\hline
chm1 &$\sigma_{227,315}$&$~~-0.10(16)$ &$~~+0.07(14)$ &$~~-0.010(12)$&&$~~~~29$& $1.1$\\
\midrule
&&&&&\\[-16pt]
&&$~~\hat l_1\times10^3$&$~~{\hat l}_2\times10^3$	&$~~L_2\times10^3$&$~~L_{68}\times10^3 $&&\\
\hline
chm2 &$\sigma_{227},\rho_{227}$&$~~+2.2(9)$&$~~-3.44(16)$&$~~+1.0(2)$&$~~+1.6(8)$&$~~~~22$&$0.9$\\
chm2&$\sigma_{315},\rho_{315}$&$~~+2.2(5)$&$~~-3.45(15)$&$~~+1.4(2)$&$~~-3(2)$&$~~~~22$&$0.9$\\
chm2&$\sigma_{227,315},\rho_{227,315}$&$~~+2.24(3)$	    &$~~-3.44(1)$    &$~~+1.2(1)$&$~~-0.1(7)$&$~~~~44$&$1.1$\\
\midrule
&&&&&\\[-16pt]
\multicolumn{2}{c}{}&$~~\hat l_1\times10^3$&$~~{\hat l}_2\times10^3$	&$~~$&$~~$&&\\
\midrule
Ref.~\cite{Guo:2016zos}&$\rho_{227,315}$&$~~+2.26(14)$&$~~-3.44(3)$&~~--&~~--&$~~~~15$&$1.3$\\
\bottomrule
\end{tabular}
\caption{Best fit parameters of considered parametrizations obtained from fits to lattice energy levels as specified in the second column. The number of data points in each fit, denoted by $n$, is stated in the seventh column. The last sub-table contains results of the [chm2] analysis of for the isovector data only, see Ref.~\cite{Guo:2016zos}.
\label{tab:fits}
}
\end{table*}

\section{Analysis of the $\pi\pi$ scattering amplitude}
\label{sec:analysis}

The extracted phase-shifts and their covariances carry the full information about the interaction of (unphysically heavy) pions at discrete values of energy (or momentum) above and even below the corresponding $\pi\pi$-threshold. From the point of view of scattering theory these points are interconnected by a function of energy (partial wave projection is assumed) which carries specific analytic properties. These properties are the guiding principle for the construction of fit functions to extract its parameters from lattice data~(as described in Section~\ref{sec:lattice} and collected at the end of this paper in Appendix~\ref{appendix:fitting}.)
Given the $S$-matrix and scattering amplitude $T$ for two-to-two scattering with $S=1-i\,T$, the unitarity constraints the imaginary part of the scattering amplitude  projected to definite isospin ($I$) and angular momentum ($L$) to be $16\pi\Im{T_{IL}^{-1}(s)}=\sqrt{1-4M_\pi^2/s}$, where $s$ denotes the square of the total four-momentum of the system. The above equation does not fix the amplitude entirely, but only up to a real-valued function  in the physical region,
\begin{align}
T_{IL}(s)=\frac{1}{K_{IL}^{-1}(s)-G(s)}\,,
\label{eq:Kmatrix}
\end{align}
where $K_{IL}(s)$ is a real-valued function, and $G(s)$ is the two pion loop-function. In dimensional regularization it reads
\begin{align}
G(s)
&
=
\frac{a(\mu)+2\ln(M_\pi/\mu)}
{16 \pi^2}
+\frac{p_{cm}}{8\pi^2\sqrt{s}}\ln\Big(\frac{2 p_{cm}+\sqrt{s}}{2p_{cm}-\sqrt{s}}\Big)\,,
\label{eq:gdimre}
\end{align}
where $p_{cm}$ is the modulus of the three-momentum in the center-of-mass system. The regularization scale and the subtraction constant are fixed throughout this study and in accordance with the discussion of Refs.~\cite{Hu:2016shf, Guo:2016zos} to {$\mu=1$~GeV} and ${a(\mu)=-1.28}$, respectively. This value results from a fit to experimental data but it can be varied in a range of $\pm 0.5$ without changing the result noticeably, i.e., the change is well absorbed into the values of the low-energy constants even if the amplitude is not explicitly scale invariant~\cite{Hu:2016shf}. Note that since $16\pi\Im G(s)=-\sqrt{1-4M_\pi^2/s}$, the above formulation of the $T$-matrix differs from the usual $K$-matrix formulation only by a re-shuffling of the $\Re G(s)$ part. The form of the function $K_{IL}(s)$ is not fixed by unitarity. In this work we will use four versions of two different types of the $K$-matrix to gauge the systematic uncertainty tied to a particular choice.

\begin{figure*}[t]
 \includegraphics[width=0.49\linewidth]{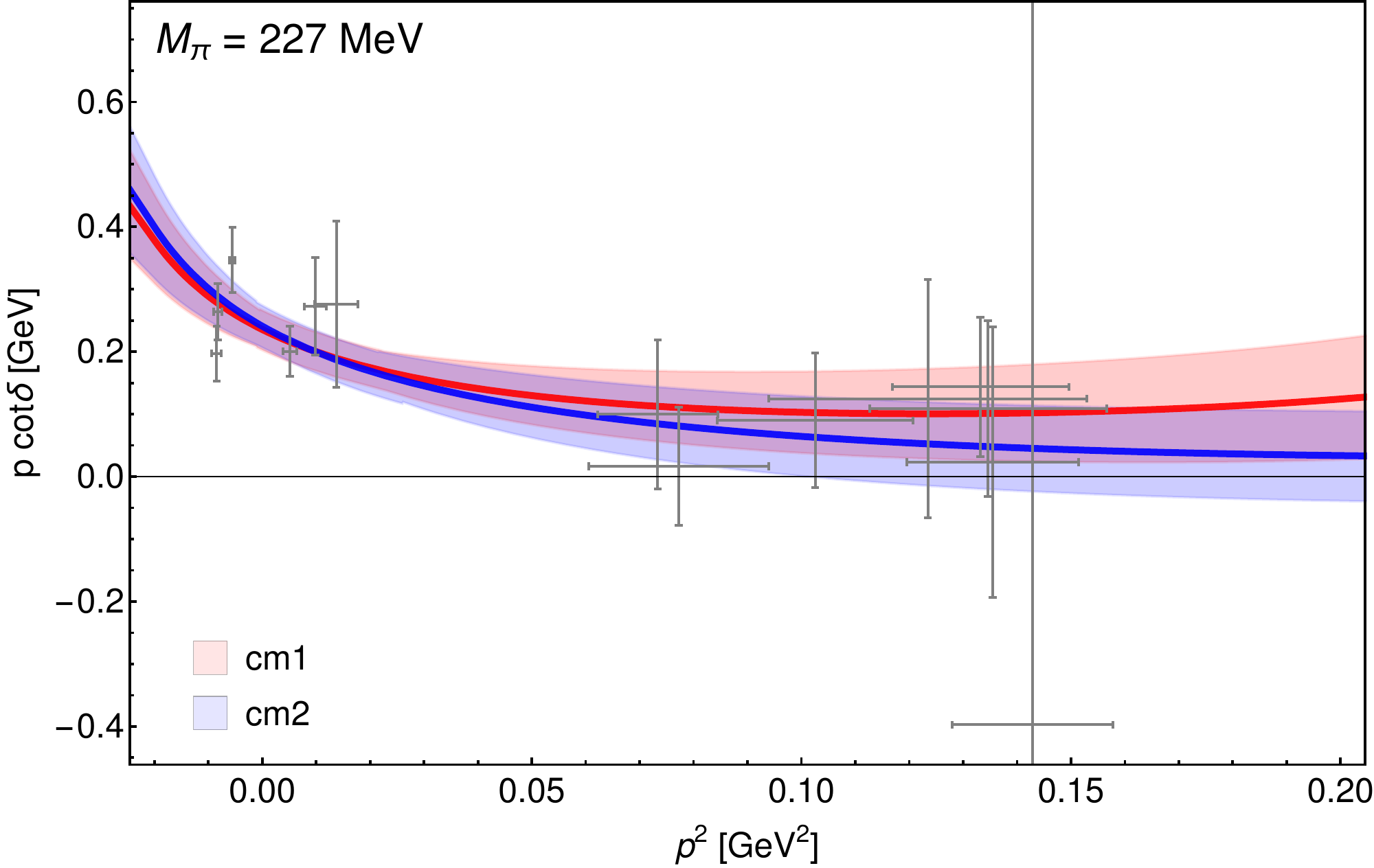}
 \includegraphics[width=0.49\linewidth]{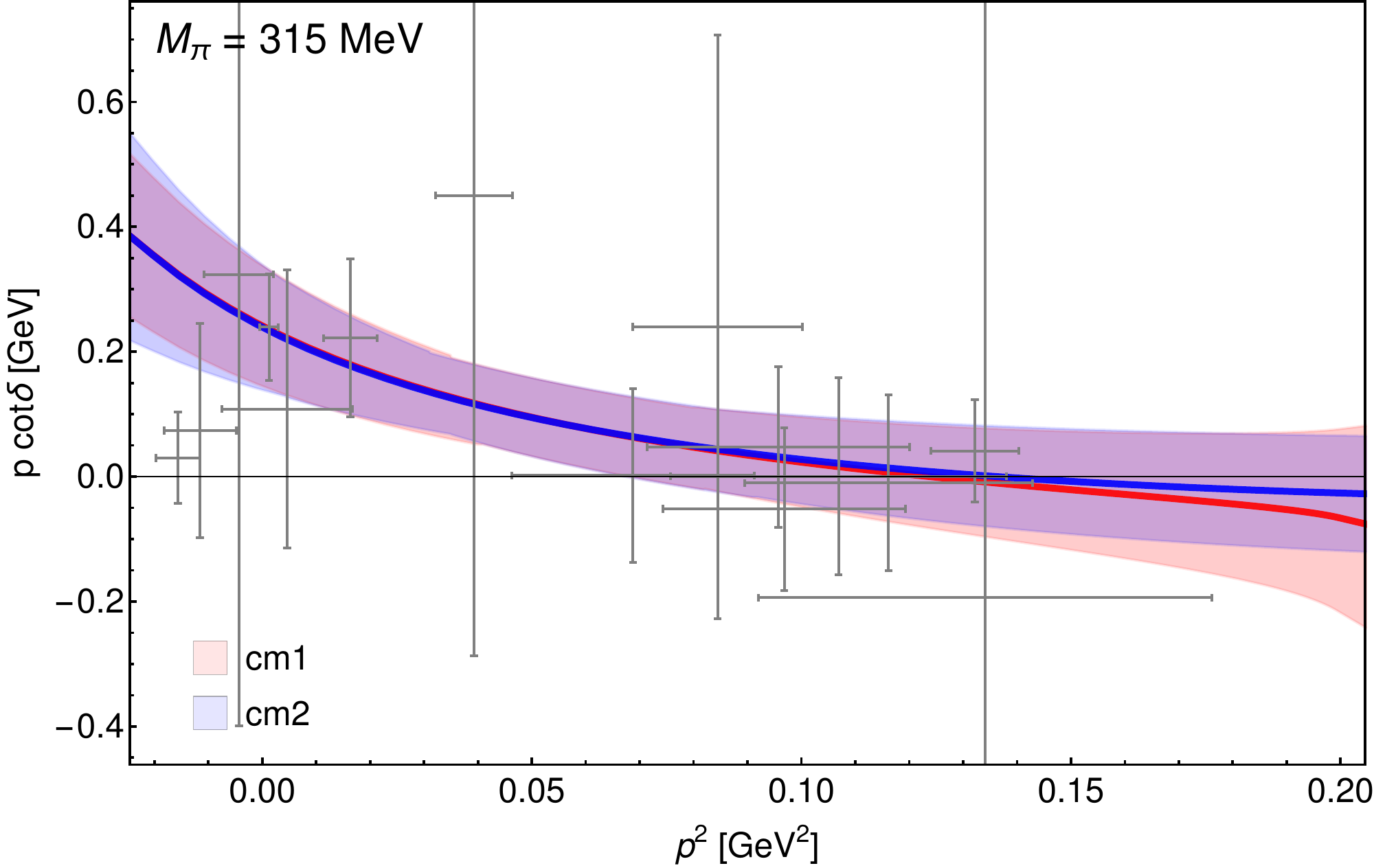}
\caption{Results of the fits to the lattice data in terms of $p\,\mathrm{cot}\,\delta$ using the conformal parametrization in two variants ([cm1] and [cm2]), cf. Eqs.~(\ref{eq:cm}) for $M_\pi=227$~MeV and $M_\pi=315$~MeV. Data points outside of the fitting region are excluded from the plot.
}
\label{fig:conformal}
\end{figure*}

For the direct use of lattice data, i.e., energy eigenvalues and their covariance matrices, we formulate the finite volume version of the scattering amplitude, i.e., of Eq.~\eqref{eq:Kmatrix}. The positions of poles of the latter give access to the discrete energy spectrum on the lattice. For boxes with asymmetry $\eta$ in the $z$ direction and in the rest frame such an amplitude can be obtained replacing $G(s)\to\tilde{G}(\eta,s)$ in Eq.~\eqref{eq:Kmatrix} with
\begin{align}
\label{eq:gtilde}
&\tilde{G}(\eta,s)=G(s)
\\&+
\lim\limits_{q_\mathrm{max}\to \infty}
\left(
\frac{1}{\eta L^3}\sum_{|{\bf q}|<q_{\mathrm{max}}}I(s,|{\bf q}|)
-
\int_{|{\bf q}|<q_{\mathrm{max}}}\frac{d^3q}{(2\pi)^3}I(s,|{\bf q}|)\right)\nonumber
\\
&\text{for~}
I(s,|{\bf q}|)=\frac{\omega_1+\omega_2}{2\omega_1\omega_2}\frac{1}{s-(\omega_1+\omega_2)^2}\,, \nonumber
\end{align}
and $\omega_i=\sqrt{|{\bf q}|^{2}+m_i^2}$, being ${\bf q}$ the momentum in the rest frame, ${\bf q}=2\pi/L\,(n_x,n_y,n_z/\eta)$. For boosts with momentum ${\bf P}$, one performs a Lorentz transformation, see Ref.~\cite{Doring:2012eu}. Note also that $\tilde{G}(\eta,s)$ is independent of $q_{\mathrm{max}}$ but does depend on the subtraction constant $a(\mu)$ via $G(s)$ as given in Eq.~\eqref{eq:gdimre}. In other words, the infinite-volume extrapolation is cut-off independent and equivalent to the L\"uscher formalism up to exponentially suppressed contributions. At the same time, the function $\tilde G$ contains a dispersive real part relevant for the parametrization of the infinite-volume amplitude itself. See Ref.~\cite{Doring:2012eu} for further details. 

The discrete energy spectrum obtained from the finite volume scattering amplitude (see Eqs.~\eqref{eq:Kmatrix} and \eqref{eq:gtilde}),
\be
K^{-1}_{00}(s_i)-\tilde G(s_i)=0 \ ,
\label{eq:levels}
\ee

depends explicitly on the form of the $K$-matrix. In the following we will specify different parametrizations for it, which in every case depend on some free parameters. These will be adjusted by minimizing   
\begin{align}
\chi^2=(\sqrt{s_i}-\sqrt{s_0})^T\cdot C^{-1}\cdot (\sqrt{s_i}-\sqrt{s}_0)\,
\label{eq:mini}
\end{align}
where $s_i$ are the fit parameter-dependent solutions of Eq.~(\ref{eq:levels}), ordered in a vector in Eq.~(\ref{eq:mini}), $\sqrt{s_0}$ indicates the vector of eigenenergies measured on the lattice, and $C$ is the covariance matrix of the correlated data. Eq.~(\ref{eq:mini}) implicitly contains a summation over contributions to the $\chi^2$ from different elongation factors $\eta$ and boosts ${\bf P}$. The covariances between data from different moving frames are taken into account.

We analyze data from different pion masses and channels individually and also simultaneously. In cases the data of the isoscalar channel (from this work) and of the $I=L=1$ $\rho$-channel~\cite{Guo:2016zos} are simultaneously fitted, Eq.~(\ref{eq:levels}) changes accordingly for the data from the $\rho$ channel, $K_{00}\to K_{11}$, and Eq.~(\ref{eq:mini}) becomes a sum over the two channels. Similarly, it becomes a sum over contributions from different pion masses in the respective simultaneous fits. Further statistical tests, such as Pearson's $\chi^2$ test will be discussed below.

\subsection{Conformal mapping}
\label{subsec:conformal}

The first type of the parametrization of the scattering amplitude relies on a general form of $K$-matrix as an analytic function of energy. As discussed in Refs.~\cite{Yndurain:2007qm,Caprini:2008fc} the convergence of a power series in energy is limited but can be improved, mapping it onto the interior of a disk limited by the right- and left-hand cuts lying on the boundary circle. We use two versions of such a mapping slightly adapted to our approach. We refer to them as [cm1] and [cm2] with the respective expansion variable 
\begin{align}
\omega^{\rm [cm1]}(s)&=\frac{\sqrt{s}-\alpha\sqrt{s_{\text{thres}}-s}}{\sqrt{s}+\alpha\sqrt{s_{\text{thres}}-s}}\,,\nonumber \\
\omega^{\rm [cm2]}(s)&=\frac{\sqrt{s}-\sqrt{c}}{\sqrt{s}+\sqrt{c}}\,,
\label{eq:cm}
\end{align}
where $\alpha$, $s_{\text{thres}}$, and $c$ are parameters of the mapping. 
In particular, $s_{\text{thres}}$ in parametrization [cm1] is the position of the next threshold opening above the $\pi\pi$ threshold. Here, we do not have a $K\bar K$ channel as in Refs.~\cite{Yndurain:2007qm,Caprini:2008fc}, but we can interpret the parameter to take account of the opening of the four-pion threshold. The [cm2] parametrization is obtained in the limit $s_{\text{thres}}\to\infty$. The quantity $\sqrt{c}$ is the expansion point in the $\sqrt{s}$ plane connected to $\alpha$ through $c=s_{\text{thres}}\alpha^2/(1+\alpha^2)$. 
For the [cm1] parametrization, $\sqrt{c}=778$~MeV (for $\alpha=1$ and $\sqrt{s_{\text{thres}}}=550\,\text{MeV}$); for [cm2], $\sqrt{c}=1$~GeV. We have checked that the results do not depend on these choices.

\begin{figure*}[t]
 \includegraphics[width=0.49\linewidth]{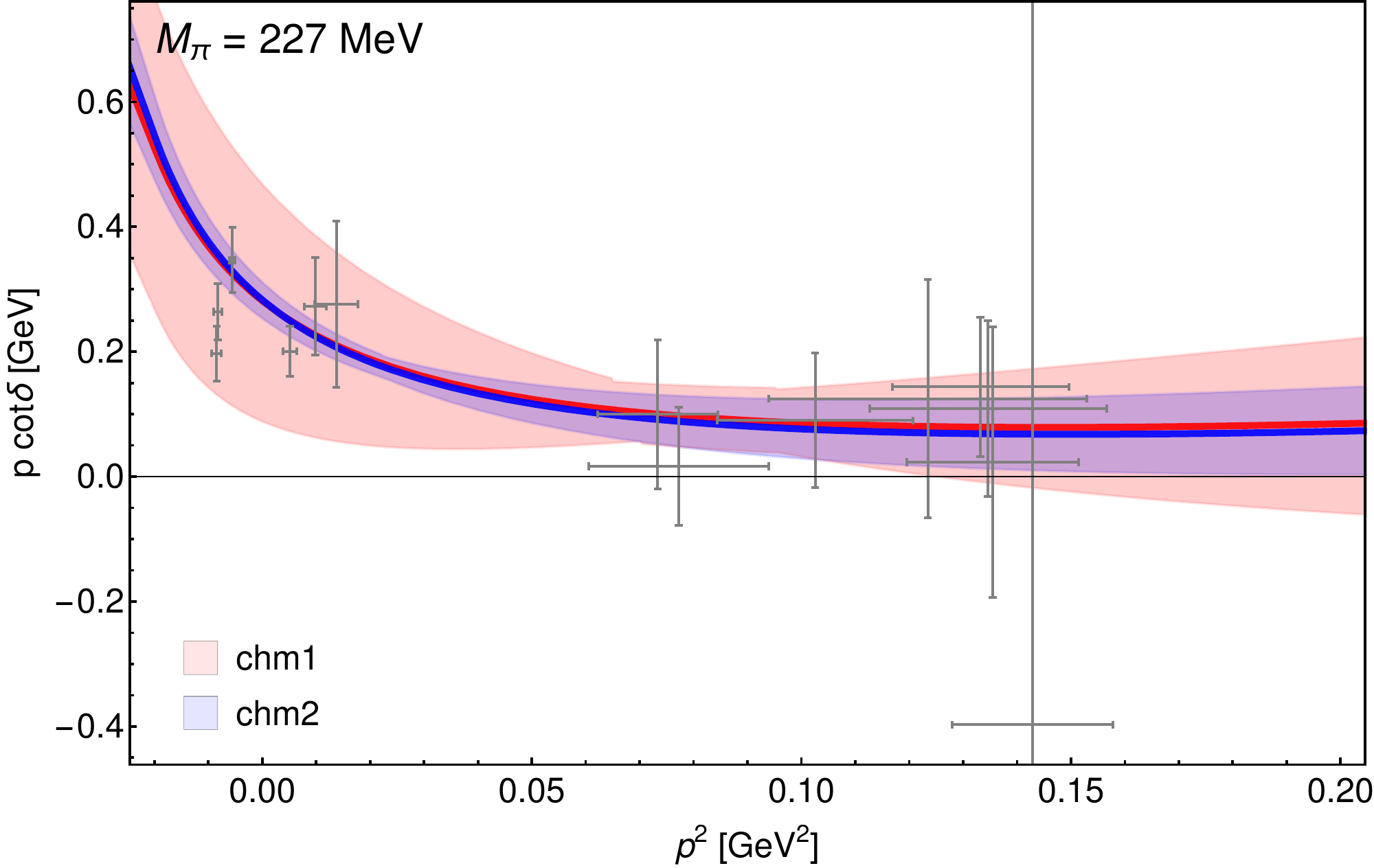}
 \includegraphics[width=0.49\linewidth]{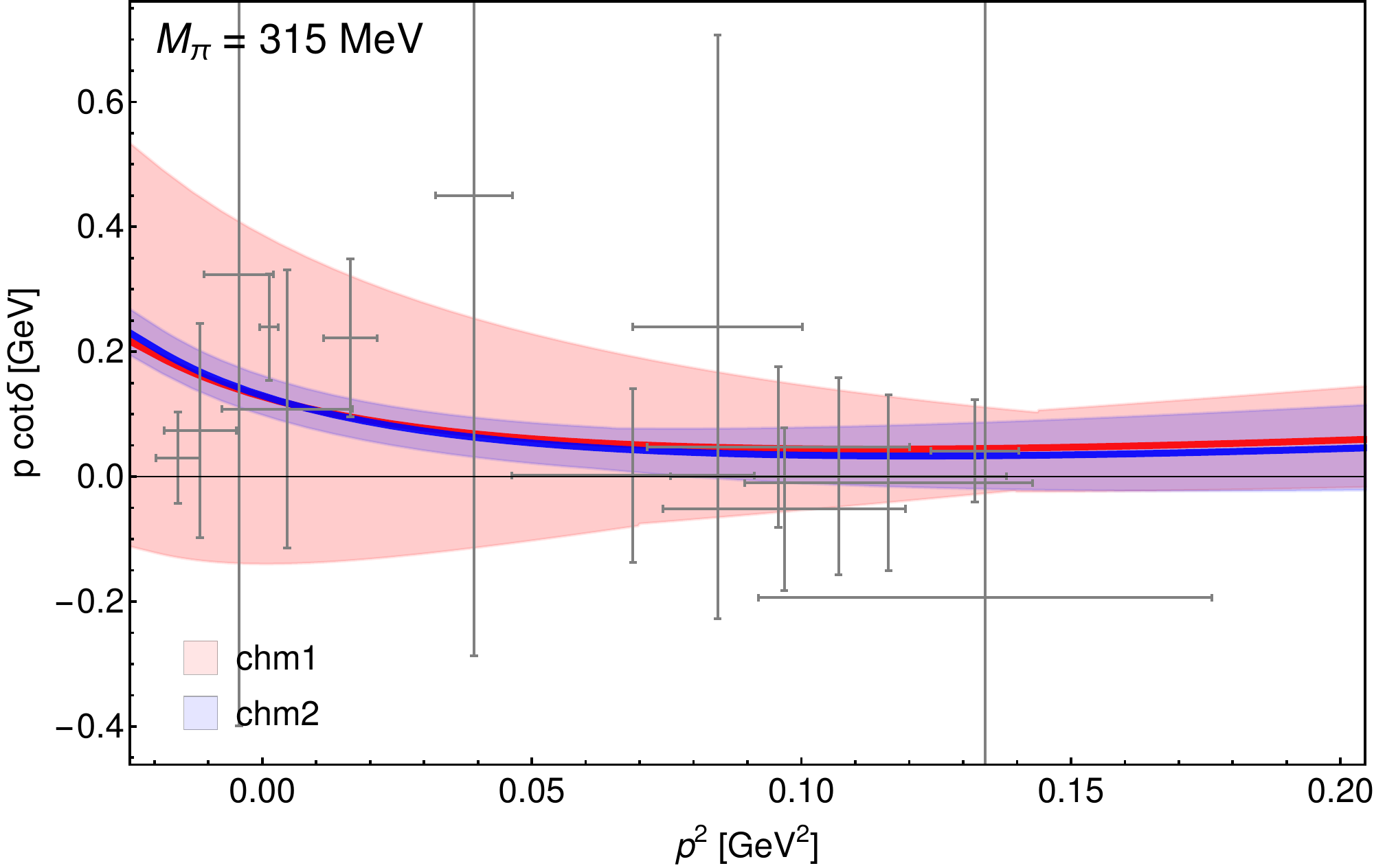}
\caption{Results of the fits to the lattice data using chiral parametrization as described in the main text. The fits are performed simultaneously to the data of both pion masses ($M_\pi=227$ and $315$ MeV), including only the data in the $\sigma$ channel [chm1], or the $\sigma$ and $\rho$ channel [chm2]. Data points outside of the fitting region are excluded from the plot.
}
\label{fig:uchpt}
\end{figure*}

For the isoscalar channel ($I=L=0$) the chiral symmetry dictates that $T_{00}(s)$ must vanish for $s= s_A\sim M_\pi^2/2$. To account for this fact the $K$-matrix in this channel takes for both versions of the mapping variable $\omega^{\rm [..]}$ the following form
\begin{align}
K_{00}^{-1}(s)=
\frac{1}{16\pi}\frac{M_\pi^2}{s_A-s}\Big(\frac{2s_A}{M_\pi\sqrt{s}}
+
B_0+B_1\omega(s)+...\Big)\,,
\end{align}
where throughout the further calculations $s_A$ is set to its leading chiral order value, i.e. $2s_A=M_\pi^2$. The number of polynomial terms in the latter equations ($\sim B_i$) is not restricted a priori. For the present case two polynomial terms ($B_0,B_1$) turn out to give sufficient flexibility in the energy variable $\omega(s)$ to fit the data. The lattice data consist of the energy eigenvalues and covariance matrices of the energy levels at two different pion masses ($M_\pi=227$ and $M_\pi=315$~MeV) in the $I=L=0$ channel which we fit individually.

Here and in section \ref{sec:uchpt} the lattice energy levels for the $\sigma$-meson are fitted up to $\sqrt{s}\approx 970$~MeV and $\sqrt{s}\approx 1070$~MeV for the light and heavy pion masses, respectively. This corresponds in both cases to $\bm p^2\approx 0.187~{\rm GeV}^2$ for the magnitude of the center of mass three-momentum $\bm p$. The data at higher $\bm p^2$ cannot be fitted by the considered parametrizations. Fitting with more flexible parametrizations would, however, require more data points in the high momentum range.

The results of the fits are collected in the first four entries of Table~\ref{tab:fits}, which all pass the Pearson's test with the total $\chi^2\approx 9$ lying inside of the 80\% confidence interval of our two-tailed test, i.e $(6,19)$ and $(7,20)$, for the light and heavy pion masses respectively. The error bands depicted in Fig.~\ref{fig:conformal} are obtained from the error ellipses of the fitting parameters. Such a procedure for the propagation of statistical uncertainties is used in the remainder of the paper. There is a clear overlap of the fits with the data. Only for the heavy pion mass and energies deep below threshold there is some discrepancy that could be significant. Note again that the phase-shifts are not fitted directly, but rather the energies extracted from lattice QCD which makes the fit results and the shown phase-shift data difficult to compare; for example, the error bars in $x$ and $y$-direction of the phase-shift data are perfectly correlated (one may think of inclined error bars), and the correlations between different phase-shift data can obviously not be visualized.

Finally, having fixed parameters of both parametrizations we perform an analytical continuation to the complex energy plane. On the second Riemann sheet of this plane we determine the position ($z_0$) and residuum ($g^2$) of the $\sigma$-resonance pole. The results are collected in the last three columns of Table~\ref{tab:poles}. They are discussed in Section~\ref{subsec:discussion} together with the results from the chiral unitary approach discussed in the next section.

\subsection{Chiral unitary approach}\label{sec:uchpt}

\begin{figure}[h!]
\includegraphics[width=1\linewidth]{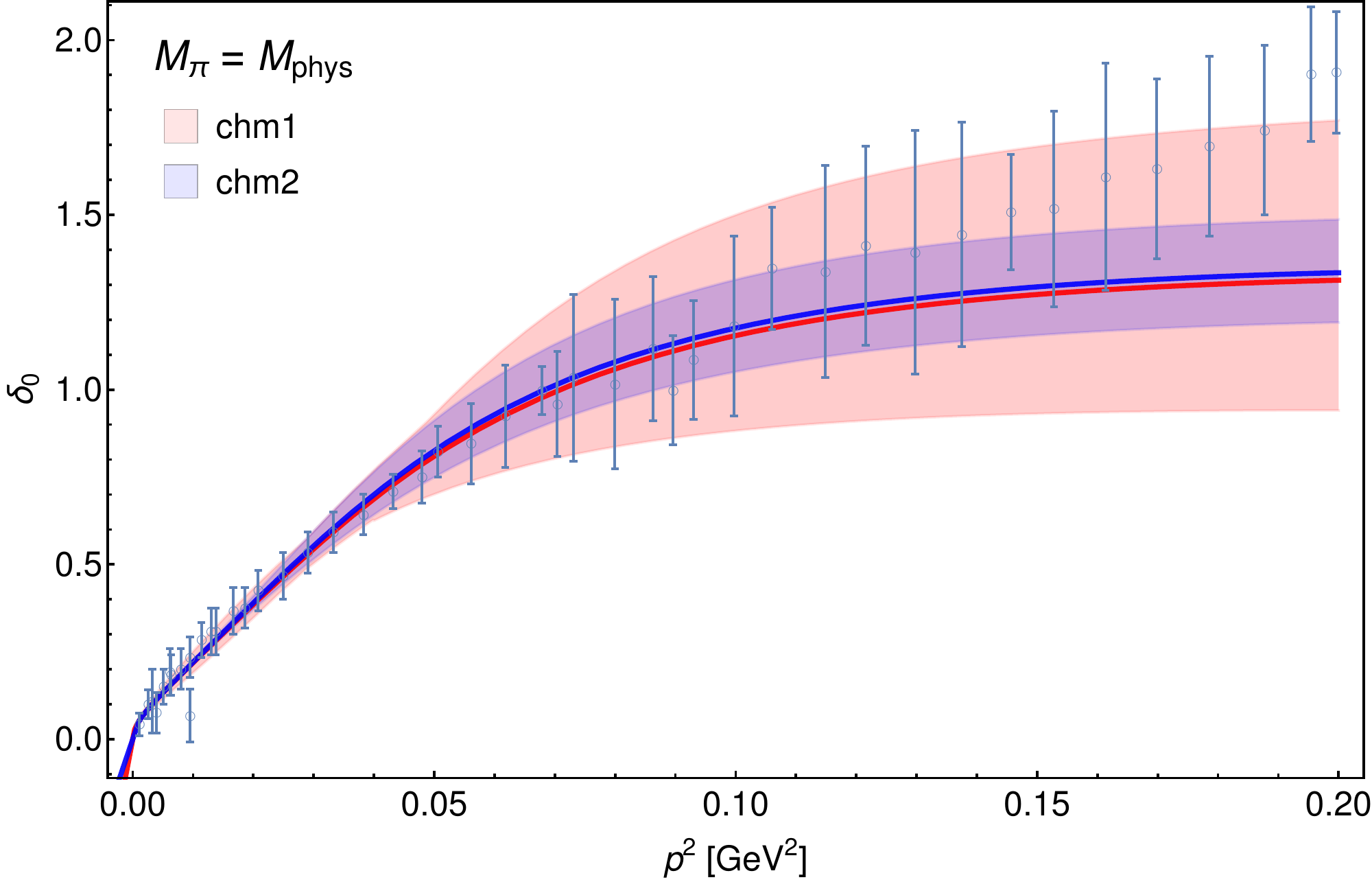}
\caption{
\label{fig:uchpt-expt}
Phase-shift extrapolated to the physical point as a result of the fits to the energy levels in the $\sigma$ (red) and the $\sigma+\rho$ (blue) channels. The experimental data is taken from Refs.~\cite{Batley:2010zza,Froggatt:1977hu,Estabrooks:1974vu,Hyams:1973zf,Protopopescu:1973sh,Grayer:1974cr}.}
\end{figure}

\begin{table*}[th]
{\renewcommand{\arraystretch}{1.7}
\begin{tabular}{lllllllllllll}
&
&\multicolumn{3}{c}{$M_\pi=138$ MeV}&
&\multicolumn{3}{c}{$M_\pi=227$ MeV}&
&\multicolumn{3}{c}{$M_\pi=315$ MeV}\\
\cline{3-5}
\cline{7-9}
\cline{11-13}
Parametrization~~&Fitted data
&~~$\Re z^*$&~~$-\Im z^*$&~~$g$&
&~~$\Re z^*$&~~$-\Im z^*$	&~~$g$&
&~~$\Re z^*$&~~$-\Im z^*$	&~~$g~~$\\    
\toprule
cm1&$\sigma_{227}$
&~~--&~~--&~~--&
&~~$460_{-60}^{+30}$&~~$180_{-30}^{+30}$&~~$3.2_{-0.1}^{+0.1}$&
&~~--&~~--&~~--\\
cm1&$\sigma_{315}$
&~~--&~~--&~~--&
&~~--&~~--&~~--&
&~~$660^{+50}_{-70}$&~~$150^{+40}_{-50}$&~~$4.0_{-0.2}^{+0.2}$\\
\midrule
cm2&$\sigma_{227}$
&~~--&~~--&~~--&
&~~$475_{-60}^{+30}$&~~$176^{+50}_{-40}$&~~$3.3_{-0.2}^{+0.3}$&
&~~--&~~--&~~--\\
cm2&$\sigma_{315}$
&~~--&~~--&~~--&
&~~--&~~--&~~--&
&~~$660_{-90}^{+50}$&~~$140^{+40}_{-50}$&~~$3.9_{-0.2}^{+0.2}$\\
\midrule
chm1&$\sigma_{227,315}$
&~~$440^{+60}_{-90}$&~~$240^{+20}_{-50}$&~~$3.0^{+0.2}_{-0.6}$&
&~~$490^{+100}_{-70}$&~~$170^{+40}_{-110}$&~~$3.0^{+0.7}_{-0.5}$&
&~~$590^{+130}_{-120}$&~~$80^{+150}_{-80}$&~~$4.0^{+4.0}_{-2.0}$\\
\midrule
chm2&$\sigma_{227}~\rho_{227}$
&~~$430^{+20}_{-30}$&~~$250^{+30}_{-30}$&~~$3.0^{+0.1}_{-0.1}$&
&~~$460^{+30}_{-40}$&~~$160^{+30}_{-30}$&~~$3.0^{+0.1}_{-0.1}$&
&~~$620^{+10}_{-80}$&~~$0^{+60}_{-0}$&~~$3.1^{+6.0}_{-3.0}$\\
chm2&$\sigma_{315}~\rho_{315}$
&~~$460^{+10}_{-15}$&~~$210^{+40}_{-30}$&~~$3.0^{+0.1}_{-0.1}$&
&~~$540^{+30}_{-40}$&~~$150^{+30}_{-30}$&~~$3.1^{+0.1}_{-0.1}$&
&~~$660^{+40}_{-60}$&~~$120^{+40}_{-40}$&~~$3.6^{+0.1}_{-0.1}$\\
chm2&$\sigma_{227,315}~\rho_{227,315}$
&~~$440^{+10}_{-16}$&~~$240^{+20}_{-20}$&~~$3.0_{-0.0}^{+0.0}$&         
&~~$500_{-20}^{+20}$&~~$160_{-15}^{+15}$&~~$3.0^{+0.0}_{-0.1}$&
&~~$600^{+30}_{-40}$&~~$80^{+20}_{-80}$&~~$3.9^{+5.0}_{-0.2}$\\
\toprule     
Ref.~\cite{Pelaez:2015qba}& experimental
&~~$449_{-16}^{+22}$&~~$275_{-12}^{+12}	$&~~$3.5^{+0.3}_{-0.2}$&
&~~--&~~--&~~--&
&~~--&~~--&~~--\\
\hline
\end{tabular}}
\caption{
\label{tab:poles}
Pole positions ($z^*$ in MeV) and corresponding couplings to the $\pi\pi$ channel ($g$ in GeV) from conformal mapping ([cm1] and [cm2]) and chiral unitary approach ([chm1] and [chm2]) as described in the main body of the manuscript. The parameters of the parametrizations are fitted to the data set specified in the second column. In the last row, the result of the analysis of experimental data~\cite{Pelaez:2015qba} is shown.}
\end{table*}

The second option for the form of the $K$-matrix explored in this work is inspired by chiral perturbation theory in two-flavor formulation. As such it contains symmetries of QCD, while it also relates different interaction channels as a full fledged quantum field theory. These two facts allow for a chiral extrapolation between different pion masses as well as for a simultaneous description of the isoscalar and isovector channels. However, the price to pay is that the full $K$-matrix would contain infinitely many terms. To make a practically feasible approach, a truncation of a chiral series is required. Different versions are in use; their theoretical properties are discussed in detail in Ref.~\cite{Bruns:2017gix}.

In the following we will use a version of chiral unitary approaches (UChPT), which, on the one hand, allows to address both ($\sigma$ and $\rho$) channels of the $\pi\pi$ scattering simultaneously, see, e.g., Refs.~\cite{Oller:1998hw,Guo:2016zos,Hu:2017wli,Hu:2016shf}. On the other hand it contains only local terms (of the leading and next-to-leading chiral order) which makes the analysis of the discrete finite volume spectrum feasible in the same way as in the case of conformal mapping, see Section~\ref{subsec:conformal}. The $K$-matrix reads in both considered channels
\begin{align}
K_{00}(s)&=\frac{3(M_\pi^2-2s)^2}{6f_\pi^2(M_\pi^2-2s)+8(L_a M_\pi^4+s(L_b M_\pi^2+L_c s))}\,,\nonumber\\\label{eq:KUChPT1}\\
K_{11}(s)&=\frac{4M_\pi^2-s}{3(f_\pi^2-8\,\hat{l}_1 M_\pi^2+4\, \hat{l}_2 s)}\,,
\label{eq:KUChPT2}
\end{align}
where $f_\pi$ is the pion decay constant fixed to its value at the given pion mass (using the data from Table~\ref{Tab:ensembles}). The low-energy constants (LECs) of the next-to-leading chiral order read~\cite{Gasser:1984gg}
\begin{align}
L_a   &=-36\hat l_1+44\hat l _2+20(5 L_2+6 L_6+3 L_8)\,,\nonumber\\
L_b   &=12\hat l_1 - 40\hat l_2 -80 L_2\,,\nonumber\\
L_c   &=11\hat l_2 +25 L_2\,,
\label{eq:lecs}\\
\hat l_2 &= 2L_1 - L_2 + L_3~~\text{and}~~
\hat l_1 = 2L_4 + L_5\nonumber\,.
\end{align}

\begin{figure*}[th]
\includegraphics[height=7.cm]{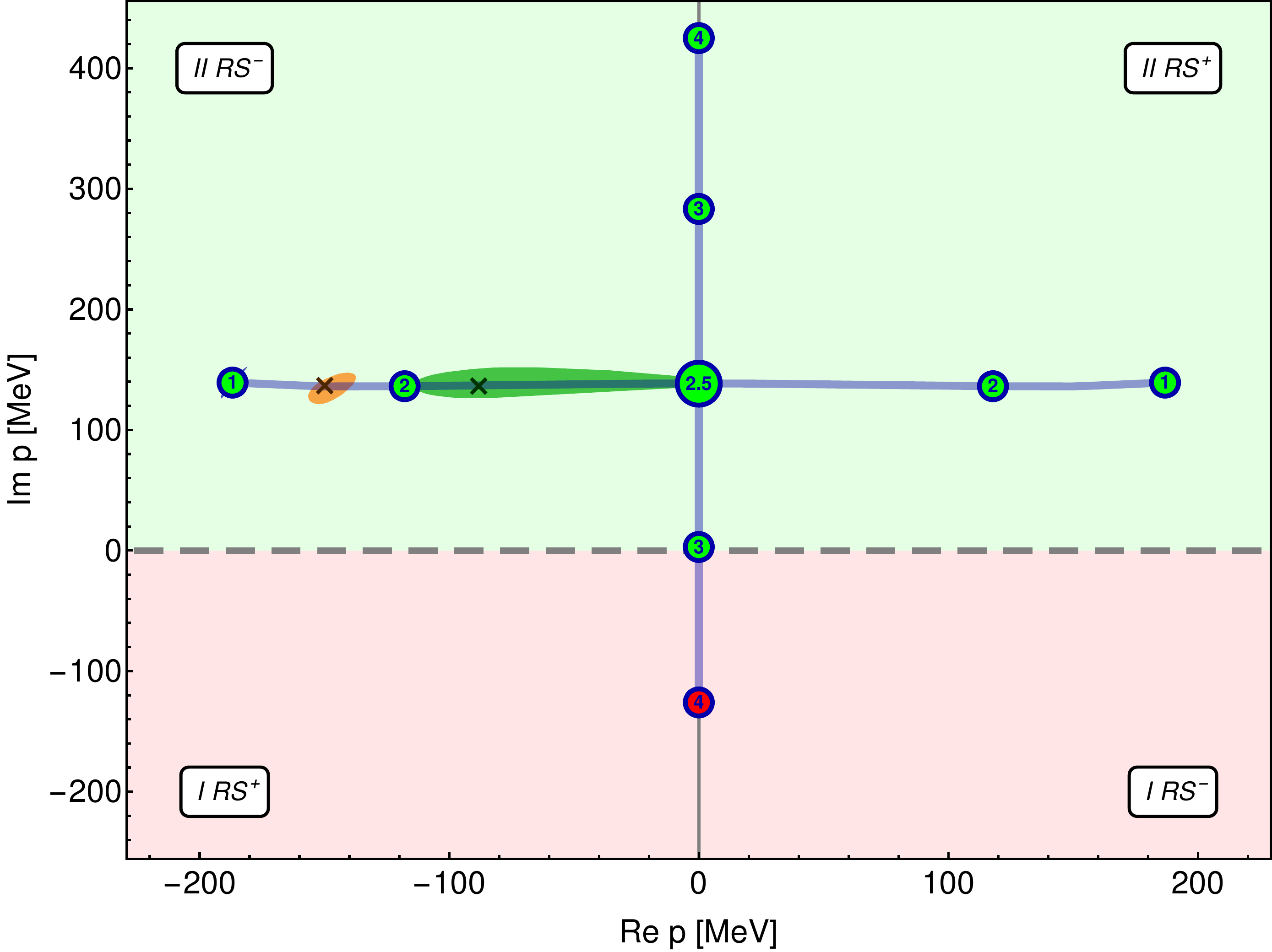}
~~
\includegraphics[height=7.cm]{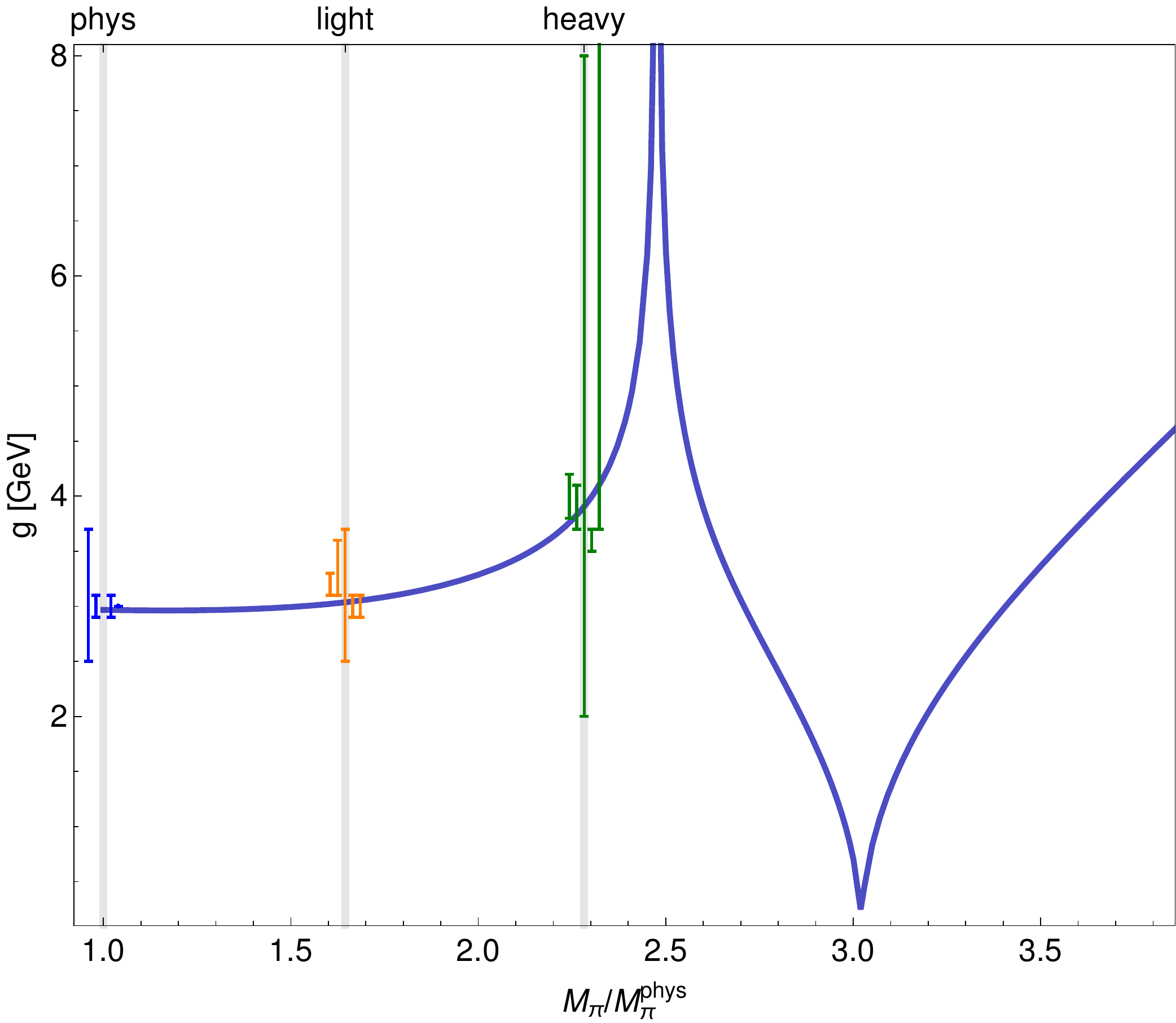}
\caption{
\label{fig:extrapolation}
Left: $M_\pi$-dependence of the pole position of the $\sigma$-resonance in the complex plane of the center of mass three-momentum from the [chm2] fit to the data from both pion masses. The dashed line represents the real   $\sqrt{s}$-axis, which connects the first ($I~RS^{\pm}$) and second ($II~RS^{\pm}$) Riemann sheets, and the subscript $+/-$ denotes the positive/negative $\sqrt{s}$ half-plane, respectively. The encircled numbers represent the pion mass in units of the physical one, while $``\mathbf{\times}"$ shows the result of the simultaneous fit to $\rho$ and $\sigma$ at light (orange) and heavy (dark green) pion masses with corresponding $1\sigma$ error areas. Right: $M_\pi$-dependence of the coupling of the $\sigma$-resonance to the $\pi\pi$ channel in the same color coding as in the left panel.
}
\end{figure*}

The model used here is identical to the one of Ref. \cite{Guo:2016zos}, where using the potential of Eq. (\ref{eq:KUChPT2}), the energy levels of the $\rho$-meson were analyzed. The same data set for the $\rho$-meson is considered here as well. Therefore, we refer to Ref. \cite{Guo:2016zos} for a detailed discussion of the results concerning the $\rho$-meson. Here, we perform a combined fit of the energy levels at the two given pion masses ($M_\pi=227$ and $315$ MeV) in the $\sigma$-resonance channel, which depends on the three combinations of LECs (fitting parameters) of Eq. (\ref{eq:KUChPT1}), $L_a$, $L_b$ and $L_c$, and combined fits of the $\sigma$ and $\rho$ channels (at one and two pion masses), being the fitting parameters in this case $\hat{l}_1$, $\hat{l}_2$, $L_2$ and $L_{68}:=L_8+2L_6$. Note that in Eq. (\ref{eq:lecs}) $L_i$ are LECs of three-flavor ChPT~\cite{Gasser:1984gg}, which, however, appear here only in four linear independent combinations corresponding to the LECs of two-flavor ChPT~\cite{Gasser:1983yg}.

In the following, we refer to [chm1] and [chm2] being the fits in the isoscalar or both isoscalar and isovector channel, respectively. The lattice data for the isovector channel includes only energy eigenvalues, corresponding to the center of mass energies $\sqrt s \in \{m_\rho-2\Gamma_\rho, m_\rho+2\Gamma_\rho\}$ where $m_\rho$, $\Gamma_\rho$ are the mass and width of the $\rho$-resonance. For the detailed discussion see Ref.~\cite{Guo:2016zos}

The best fit parameters are collected in Table~\ref{tab:fits}, see the entries ``chm1 $\sigma_{227,315}$'' and ``chm2 $\sigma_{227,315}\rho_{227,315}$''. These fits pass the two-tailed Pearson's test with  $\chi^2\approx29$ for [chm1], and $\chi^2\approx44$ for [chm2], which both are inside of the corresponding 80\% intervals of $(17,36)$ and $(29,52)$, respectively. The corresponding phase-shifts are depicted in Fig.~\ref{fig:uchpt}. 

For both best fits (fitting in both cases $M_\pi=227$ and $315$ MeV simultaneously) we perform an extrapolation to the physical point, predicting the phase-shifts and comparing them with the experimental data from Refs.~\cite{Batley:2010zza,Froggatt:1977hu,Estabrooks:1974vu,Hyams:1973zf,Protopopescu:1973sh,Grayer:1974cr}. The result is depicted in Fig.~\ref{fig:uchpt-expt}, and shows a good agreement with the experimental data below energies of $950$ MeV. Some deviation from the data becomes evident at higher energies, associated with the presence of $f_0(980)$ not captured by the two-flavor parametrization of the $K$-matrix. Furthermore, we perform an analytic continuation to the second Riemann sheet of the complex energy plane, determining the pole position and the coupling to the $\pi\pi$ channel. The results including a prediction at the physical point are collected in the rows 5-8 of Table~\ref{tab:poles}. In principle, the chirally inspired approach used here allows for a prediction of the full chiral trajectory of the pole positions and their residua, which will be discussed in the next section.

In tables~\ref{tab:fits} and \ref{tab:poles} also other fits are quoted. In particular, the chiral unitary approach has been fitted to the lighter pion mass alone (``$\sigma_{227},\,\rho_{227}$'') and to the heavier pion mass alone (``$\sigma_{315},\,\rho_{315}$''). For the low-energy constants we observe agreement between these cases, and also between these cases and the combined fit to both masses. A similar agreement is observed for the pole positions and residues which shows that the data are consistent under the fit hypothesis and, independently of the pion masses in the lattice calculation, lead to similar predictions for the $\sigma$ properties at physical pion masses. 

\subsection{Discussion of the results}
\label{subsec:discussion}

Both types of parametrization of the scattering amplitude have various advantages, complementing each other. On the one hand, parametrization on the basis of chiral amplitudes up to the next-to-leading order allows to perform an extrapolation of the results in pion masses ([chm1]), describing at the same time isoscalar and isovector channels of $\pi\pi$ scattering ([chm2]). On the other hand, the parametrization based on the conformal mapping of the expansion variable ([cm1] and [cm2]) yield a more model-independent form of the amplitude.  

As demonstrated in Figs.~\ref{fig:conformal} and \ref{fig:uchpt} all considered parametrizations lead to a good agreement with the fitted lattice data. Also the experimental data from Refs.~\cite{Estabrooks:1974vu,Batley:2010zza,Hyams:1973zf,Protopopescu:1973sh,Grayer:1974cr,Gunter:1996ij} lies in the 1$\sigma$ error band around the extrapolated phase-shift from both parameterizations [chm1] and [chm2].

The analytical continuation of the scattering amplitude to the second Riemann sheet reveals the presence of a pole corresponding to the $\sigma$-resonance. The real and imaginary parts of the pole position as well as the residuum  give the information about the mass, width and coupling to the $\pi\pi$ channel of this exited state, respectively. The results are collected in Table~\ref{tab:poles}. For comparison, we also quote there the result of the dispersion analysis of the experimental data from Ref.~\cite{Pelaez:2015qba}, which overlaps (within 1-2 $\sigma$ uncertainty) with our extrapolations  [chm1] and [chm2]. Additionally, the comparison between results of [cm1]/[cm2] and [chm1]/[chm2] shows a good agreement of both approaches for the lower pion mass ($M_\pi=227\MeV$). For the higher pion mass, the pole position in the individual [chm2] fit for this mass is in good agreement with the result of the conformal parameterizations. In the chiral fits to both pion masses combined, ``[chm1], $\sigma_{227},\,\sigma_{315}$'', the data from the lighter pion mass pushes the state towards the real axis making it, for some value in the uncertainty area, a virtual bound state. As a result, the pole position and residue have large uncertainties (see Fig. \ref{fig:extrapolation} and explanation below). Overall, the systematic uncertainty tied to the use of one or another parametrization appears to be smaller than the statistical one. 

With this in mind we make a prediction of the $\sigma$ pole position and the corresponding coupling to the $\pi\pi$ channel as a continuous function of the pion mass based on [chm2] fitted to both sets of lattice data, $\sigma$ and $\rho$, and both pion masses simultaneously. To remind the reader these sets are obtained from calculations at $M_\pi\approx 1.65\,M_\pi^{\rm phys}$ and $M_\pi\approx 2.3\,M_\pi^{\rm phys}$. The result of the extrapolation is depicted in Fig.~\ref{fig:extrapolation} and exhibits three major regions: 
\begin{enumerate}[leftmargin=0.5cm]
\item With pion mass increasing from the physical one, the $\sigma$-resonance  becomes lighter in units of pion mass, but couples more strongly to the $\pi\pi$ channel. The same happens with the reflected pole in the positive half-plane of the second Riemann sheet. At $M_\pi\approx 2.5\,M_\pi^{\text{phys}}$ both poles meet at the real energy axis below threshold on the second Riemann sheet (i.e. w.r.t. the cms-momentum - positive imaginary half-axis), becoming virtual bound states. 
\item With higher pion mass, both poles evolve on the real axis towards and away from the $\pi\pi$ threshold ($p=0$), respectively. At $M_\pi\approx 3 M_\pi^{\text{phys}}$ one pole reaches the two-pion threshold, where the coupling $g$ vanishes.
\item At higher pion masses the $\sigma$-resonance becomes a bound state, thus appearing as a pole on the first Riemann sheet below threshold or negative imaginary half-axis in the complex cms-momentum plane. The coupling to two pions increases in this region monotonically.   
\end{enumerate}

The behavior described above has been pointed out first in Ref.~\cite{Hanhart:2008mx} using experimental data only and later in Ref.~\cite{Doring:2016bdr} with the input from lattice data of Refs.~\cite{Dudek:2012xn,Wilson:2015dqa,Briceno:2016mjc}. The present study  favors a bound $\sigma$ state at a pion mass of around $M_\pi\approx 3 M_\pi^{\text{phys}}$. However, we note that the uncertainties grow rapidly with increasing pion mass as can be seen in the size of green $1\sigma$ error area on pole positions and error bars on $g$ at heavy pion mass (see Fig.~\ref{fig:extrapolation}).

From Table~\ref{tab:poles} we note that the un-extrapolated results of conformal parametrizations ([cm1] and [cm2]) indicate that the systematical uncertainty tied to the choice of the  ($K$-matrix) parametrization and approximation of the left hand cut is well under control, at least at the present level of statistical uncertainty of the lattice data. We have estimated that systematical uncertainty due to fitting window and lattice spacing in the calculation of the lattice data may have higher importance and lead to several percent uncertainty for the mass and width of the $\sigma$-resonance.

\section{Summary}
\label{sec:summary}

We performed a calculation of the phase-shifts in the isoscalar/scalar $\pi\pi$ channel in the elastic region for two quark masses corresponding to $M_\pi=227\MeV$ and $315\MeV$. For each quark mass we used ensembles with three different volumes to help us determine the phase-shifts in different kinematic regions.

To extract the parameters of the $\sigma$-resonance, we have to use a parametrization that satisfies physical constraints, in particular unitarity, analyticity, and proper chiral behavior. To gauge the systematics associated with the choice of such parametrizations, we used two types of approaches (each in several variants): a generic one that makes no assumption about the underlying dynamics, and a chiral perturbation theory inspired one that allows to extrapolate the resonance parameters to different (i.e., physical) pion mass.

The systematic errors associated with the choice of parametrization are about 10\% for the pole position. The other sources of systematic errors that we assessed are the discretization errors and the fit-window for the extraction of the lattice QCD energies. The discretization errors are estimated based on the spread of lattice spacing results determined using various methods to calibrate it; we estimated this to be around 2-3\%~(see \cite{Guo:2016zos} for more details). For the fit window systematics, we computed the pole position based on energies extracted using slightly shifted fit windows; the shift on the pole position was at the level of 1\%.

One of the strengths of the chiral parametrization is that it allows us to fit simultaneously both the $\sigma$ and $\rho$ channel, for both pion masses. We find that the model describes the data well and that the results extracted from the simultaneous fit to both channels agree well with the $\sigma$-channel fit results. We use the combined fit to extrapolate to the physical point and, based on the position of the pole in the complex energy plane, we find that ${M_\sigma = (440^{+10}_{-16}(50)-i\,240(20)(25))\MeV}$. Here the first error is the stochastic error and the second one is the combined systematic error discussed above.

The extrapolation to the physical point agrees with the 
experimental phase-shifts and the pole mass and width of the $\sigma$ is compatible with the result of recent analyses based on experimental data.

\bigskip

\begin{acknowledgments}

D.G. and A.A. are supported in part by the National Science Foundation CAREER grant PHY-1151648 and by U.S. DOE Grant No.DE-FG02-95ER40907. R.M. acknowledges financial support from the Fundaç\~ao de amparo \`a pesquisa do estado de S\~ao Paulo (FAPESP). M.M. is thankful to the German Research Foundation (DFG) for the financial support, under the fellowship MA 7156/1-1, as well as to the George Washington University for the hospitality and inspiring environment. A.A. gratefully acknowledges the hospitality of the Physics Department at the University of Maryland where part of this work was carried out. M.D. acknowledges support by the National Science Foundation (CAREER grant no. PHY-1452055) and by the U.S. Department of Energy, Office of Science, Office of Nuclear Physics under contract number DE-AC05-06OR23177. The computations were carried out on the GWU Colonial One computer cluster and the GWU IMPACT collaboration clusters; we are grateful for their support.

\end{acknowledgments}

\bibliography{GAMMDref}

\appendix
\begin{widetext}
\newpage

\section{Extracted energies and correlation matrices}
\label{appendix:fitting}
In this section we list the fit result and the corresponding fit quality for each energy level and for each ensemble in Table~\ref{Tab:fit_details_sigma}. In this table, $d$ stands for the time interval of shifted correlator. In the rest frame ($\bm P = (0,0,0)$), we need to have $d\neq 0$ to subtract the vacuum contribution. In the boost frame where the total momentum in our study is $\bm P = (0,0,1)$, we use the normal correlator to extract the energy spectrum. Therefore, we set the $d$ to zero in this case. $Q$ represents the confidence level of the fit, that is the probability
under ideal conditions that the $\chi^2$ is larger than the fit result.

\begin{table*}[b]
\begin{tabular}{@{}*{6}{>{$}c<{$}}*{2}{>{\hspace{3mm}}l!{\hspace{3mm}}}*{1}{>{$0.}c<{$}}@{}}
\multicolumn{9}{c}{$M_\pi=315\MeV$}\\
\toprule
 \mathbf{P} & \eta  &n & d & t_0 & \text{fit window} & \multicolumn{1}{c}{aE} & \multicolumn{1}{c}{$\chi^2$/dof} & \multicolumn{1}{c}{Q}\\
\midrule
(0,0,0) & 1.0   & 1 & 2 & 2 & 3-13 & 0.355(9) & 0.99 & 44\\
                   &         & 2 & 2 & 2 & 2-8 & 0.54(3) & 0.89 & 47\\
                   &         & 3 & 2 & 2 & 2-7 & 0.66(2) & 0.91 & 44\\
                   & 1.25 & 1 & 2 & 2 & 3-12 & 0.363(15) & 0.92 & 49\\
                   &         & 2 & 2 & 2 & 2-8 & 0.50(2) & 0.88 & 47\\
                   &         & 3 & 2 & 2 & 2-9 & 0.59(2) & 0.86 & 48\\
                   & 2.0   & 1 & 2 & 2 & 4-21 & 0.378(7) & 1.02 & 43\\
                   &         & 2 & 2 & 2 & 2-7 & 0.457(8) & 1.05 & 37\\
                   &         & 3 & 2 & 2 & 3-9 & 0.54(2) & 0.84 & 50\\
\midrule
     (0,0,1) & 1.0     & 1 & 0 & 2 & 2-9 & 0.492(5) & 0.27 & 93\\
                 &          & 2 & 0 & 2 & 2-7 & 0.693(3) & 0.46 & 71\\
                 & 1.25  & 1 & 0 & 2 & 4-13 & 0.447(16) & 1.4 & 20\\
                 &          & 2 & 0 & 2 & 2-8 & 0.60(3) & 0.65 & 62\\
                 & 2.0    & 1 & 0 & 2 & 4-21 & 0.410(2) & 0.70 & 77\\
                 &          & 2 & 0 & 2 & 4-8 & 0.54(2) & 0.92 & 40\\
                 &          & 3 & 0 & 2 & 2-6 & 0.60(1) & 0.07 & 93\\
\bottomrule
\end{tabular}
~~~~~~
\begin{tabular}{@{}*{6}{>{$}c<{$}}*{2}{>{\hspace{3mm}}l!{\hspace{3mm}}}*{1}{>{$0.}c<{$}}@{}}
\multicolumn{9}{c}{$M_\pi=227\MeV$}\\
\toprule
\mathbf{P} & \eta  &n & d & t_0 & \text{fit window} & \multicolumn{1}{c}{aE} & \multicolumn{1}{c}{$\chi^2$/dof} & \multicolumn{1}{c}{Q}\\
\midrule  
 (0,0,0) & 1.0   & 1 & 3 & 3 & 3-14 & 0.256(2) & 0.97 & 46\\
             &       & 2 & 3 & 3 & 2-9 & 0.48(3) & 0.98 & 43\\
             &       & 3 & 3 & 3 & 3-11  & 0.60(2)   & 1.03 & 40\\
             & 1.17  & 1 & 3 & 3 & 4-13 & 0.256(3)   & 0.23 & 98\\
             &       & 2 & 3 & 3 & 2-8 & 0.44(3)   & 0.61 & 65\\
             &       & 3 & 3 & 3 & 2-6 & 0.54(2)  & 0.37 & 69\\
             & 1.33  & 1 & 3 & 3 & 2-15 & 0.264(2) & 0.77 & 68\\
             &       & 2 & 3 & 3 & 2-14 & 0.44(1) & 0.28 & 99\\
             &       & 3 & 3 & 3 & 2-8 & 0.53(2)    & 0.78 & 54\\
\midrule
     (0,0,1) & 1.0   & 1 & 0  & 2 & 5-15 & 0.409(7) & 0.09 & 91\\
                &         & 2 & 0  & 2 & 2-8 & 0.59(3) & 0.81 & 51\\
                & 1.17  & 1 & 0  & 2 & 6-17 & 0.379(5)   & 0.81 & 60\\
                &          & 2 & 0  & 2 & 3-8 & 0.575(16)   & 1.08 & 36\\
                & 1.33  & 1 & 0 & 2 & 4-15 & 0.353(3) & 0.38 & 94\\
                &         & 2 & 0  & 2 & 3-11 & 0.55(3) & 0.38 &94\\
                &         & 3 & 0  & 2 & 3-9 & 0.648(4)    & 1.08 & 36\\
\bottomrule
\end{tabular}
\caption{Extracted energies and fitting details for $\sigma$-meson. $\eta$ is the elongation factor. $t_0$ is the initial time we choose for the variational analysis. $d$ is the shift interval we used in the shift correlator method.}
\label{Tab:fit_details_sigma}
\end{table*}

To determine resonance parameters
by fitting a functional description to our phase-shifts we need to take into account cross-correlations between the extracted energies. 
The energies extracted from different
ensembles are uncorrelated, but there will be correlations between the energy levels extracted from the same ensemble including the correlations between the energy levels in the rest frame and the boosted frame. We computed these covariance matrices using a jackknife resampling procedure.
These matrices are listed in Table~\ref{Tab:cov_sigma}. In the
left column from top down we list the ensembles ${\cal E}_{1,2,3}$ corresponding to $m_\pi\approx315\MeV$
and on the right ${\cal E}_{4,5,6}$ corresponding to $m_\pi\approx227\MeV$. The
order of the levels in each matrix corresponds to the order they appear in Table~\ref{Tab:fit_details_sigma}.
\begin{table*}[tbh!]
\begin{tabular}{*{2}{>{$}l<{$}}}
\left(
\begin{array}{ccccc}
75.1  & -29.4 & -7.2   & -2.37     & -7.14\\
         & 1137 & 653.6 & -23.2    & 63.6\\
         &          & 1345  & -6.37    &30.3\\
         &          &           & 57.6    & 116.5\\
         &          &           &         & 1025.1\\                 
\end{array}
\right) 
\times10^{-6}
&
\left(
\begin{array}{ccccc}
5.19  & -7.5 & 5.97   & 1.29     & -4.27\\
         & 806.6 & 377.4 & -5.84    & -60.0\\
         &          & 1068  & -7.43    &-44.4\\
         &          &           & 56.5    & -37.9\\
         &          &           &         & 790\\                 
\end{array}
\right)\times10^{-6}
\\
\left(\begin{array}{ccccc}
193  &  -37.2 & -56.3   & 16.13 & -7.89\\
       &   1130 & 1253   & -11.1  &-67.4\\
       &           & 2848  &-8.55  & 61.8\\
       &           &          & 414.0    & 265.5\\
       &           &          &           & 1552\\
\end{array}
\right)\times10^{-6} 
&
\left(
\begin{array}{ccccc}
7.06  & -2.64 & 0.601   & 0.561     & 1.65\\
         & 820.5 & 403.6 & -2.34    & -49.2\\
         &          & 432  & 3.13    &-38.6\\
         &          &           & 17.5    & -12.8\\
         &          &           &         & 465.5\\                 
\end{array}
\right)\times10^{-6}
\\
\left(
\begin{array}{cccccc}
163.8   & -70.4  & 156.6  & -1.003   & -2.66   & -5.65 \\
         & 137.6   &117.6   &1.57 & -0.846 & -0.682\\
         &          &  966.1 & -0.226   & 30.8   & -29.6 \\
         &          &         &  10.4   & 14.6   &2.95\\
         &          &         &            & 479.6   & -51.8\\
         &          &         &            &           & 104.2\\                 
\end{array}
\right)\times10^{-6}
&
\left(
\begin{array}{cccccc}
2.31   & 5.24  & 1.23  & 0.868   & 1.91   & -0.118 \\
         & 373.5   &263.8   &3.78 & -38.1 & -2.20\\
         &          &  513.2 & 0.927   & -16.04   & -3.33 \\
         &          &         &  7.63   & 4.31   &1.39\\
         &          &         &            & 1648   & 26.04\\
         &          &         &            &           & 15.6\\                 
\end{array}
\right)\times10^{-6}
\end{tabular}
\caption{Covariance matrices for the energies extracted from each ensembles. In the
left column from top down we list the ensembles ${\cal E}_{1,2,3}$ corresponding to $m_\pi\approx315\MeV$
and on the right ${\cal E}_{4,5,6}$ corresponding to $m_\pi\approx227\MeV$. The
order of the levels in each matrix corresponds to the order they appear in Table~\ref{Tab:fit_details_sigma}.} 
\label{Tab:cov_sigma}
\end{table*}

\end{widetext}


\end{document}